\documentclass{aastex631}

\usepackage{amsmath}
\usepackage{mathtools} 
\usepackage[T1]{fontenc} 

\allowdisplaybreaks

\renewcommand{\dh}{\partial}
\renewcommand{\d}{\mathrm{d}}
\newcommand{\meanBr}[1]{\left<#1\right>}
\newcommand{\curl}{\nabla\!\times\!}
\newcommand{\defn}{\equiv}
\newcommand{\abs}[1]{\left|#1\right|}
\newcommand{\op}[1]{\operatorname{#1}}
\newcommand{\bigO}{\mathcal{O}}
\newcommand{\scalesAs}{\sim}

\renewcommand{\vec}[1]{\boldsymbol{#1}}
\newcommand{\Heaviside}{\operatorname{\text{$\Theta$}}}
\newcommand{\cross}{\times}
\newcommand{\Dirac}{\operatorname{\delta}}
\newcommand{\Rm}{\ensuremath{\operatorname{Rm}}} 
\newcommand{\PrM}{\ensuremath{\operatorname{Pr}_\mathrm{m}}} 
\newcommand{\St}{\ensuremath{\operatorname{St}}} 
\newcommand{\exchangeArrow}{\leftrightarrow}
\newcommand{\continuedTerm}{\phantom{===}}

\newcommand{\FT}[1]{\widetilde{#1}} 
\newcommand{\mcB}{\mathcal{B}} 
\newcommand{\tcf}{\mathfrak{D}} 
\newcommand{\iDelLam}{\mathcal{A}} 
\newcommand{\ssdShTerm}[1]{\left[\mathcal{I}^{#1}\right]} 
\newcommand{\sym}{\Upsilon} 
\newcommand{\tauscl}{\bar{\tau}} 
\newcommand{\etascl}{\bar{\eta}} 

\newcommand{\Textcite}{\Citet}
\newcommand{\parencite}{\citep}

\shorttitle{SSD with nonzero correlation time}
\shortauthors{Gopalakrishnan and Singh}

\begin{document}

\title{Small-scale dynamo with nonzero correlation time}
\correspondingauthor{Kishore Gopalakrishnan}

\author[0000-0003-2620-790X]{Kishore Gopalakrishnan}
\affiliation{IUCAA, Post Bag 4, Ganeshkhind, Pune 411007, India}
\email{kishoreg@iucaa.in}

\author[0000-0001-6097-688X]{Nishant K Singh}
\affiliation{IUCAA, Post Bag 4, Ganeshkhind, Pune 411007, India}
\email{nishant@iucaa.in}

\begin{abstract}
	The small-scale dynamo is typically studied by assuming that the correlation time of the velocity field is zero.
	Some authors have used a smooth renovating flow model to study how the properties of the dynamo are affected by the correlation time being nonzero.
	Here, we assume the velocity is an incompressible Gaussian random field (which need not be smooth), and derive the lowest-order corrections to the evolution equation for the two-point correlation of the magnetic field in Fourier space.
	Using this, we obtain the evolution equation for the longitudinal correlation function of the magnetic field
	($M_L$)
	in nonhelical turbulence, valid for arbitrary Prandtl number.
	The non-resistive terms of this equation do not contain spatial derivatives of $M_L$ of order greater than two.
	We further simplify this equation in the limit of high Prandtl number, and find that the growth rate of the magnetic energy is much smaller than previously reported.
	Nevertheless, the magnetic power spectrum still retains the Kazantsev form at high Prandtl number.
\end{abstract}

\keywords{%
Magnetohydrodynamics (1964);
Astrophysical magnetism (102);
Perturbation methods (1215).
}

\section{Introduction}
Magnetic fields are ubiquitous in astrophysics, being found in stars, planets, galaxies, and even galaxy clusters \citep[for a review, see][section 2]{KanduPhysicsReports2005}.
Dynamo theory attempts to explain the generation and sustenance of such magnetic fields \citep{MoffattMagFieldGenBook, KrauseRadler80, KanduPhysicsReports2005, ShukurovKanduBook}.
Typically, magnetic fields ordered on the system scale are explained by appealing to mean-field dynamo theory, which suggests that fluid motions correlated at some scale can generate magnetic fields ordered on much larger scales (the `large-scale dynamo' or LSD).
However, it is well known that in a turbulent fluid, intermittent magnetic fields can be generated which are typically ordered on scales comparable to or smaller than that of the velocity field (the `small-scale dynamo' or SSD) \citep{kazantsev1968enhancement, Molchanov85}.
The SSD grows on a timescale comparable to the eddy turnover time; this is much smaller than the timescale for growth of the LSD and the typical ages of astrophysical objects.
\Citet{kulsrud92} argue that the presence of SSD-generated magnetic fields invalidates the usual treatment of the LSD.
While their conclusion has since been challenged \citep[e.g.][]{Sub98}, magnetic
fields generated by the SSD are still expected to affect the evolution of any object that contains a sufficiently turbulent plasma (i.e.\@ where the magnetic Reynolds number, $\Rm$, is above some critical value).

In general, the Lorentz force turns the evolution of the magnetic field into a nonlinear problem, which is difficult to study analytically.
As a first step, one can study the kinematic limit, where the magnetic field is assumed to be so weak that the effect of the Lorentz force on the velocity field can be neglected.
The statistical properties of the velocity field can then be treated as given quantities, and we are interested in the statistical properties of the magnetic field.

Since the evolution equation for the moment of a particular order of the magnetic field involves mixed higher-order moments of the velocity and magnetic fields, one ends up with a hierarchy of coupled evolution equations for the moments.
One needs to make additional assumptions in order to truncate this hierarchy (closure).

The standard treatment \citep{kazantsev1968enhancement, vainshtein86, ScBoKu02} is to model the velocity field as a Gaussian random field, such that all its higher moments can be expressed in terms of the first two moments.\footnote{
\Citet{Sub97}, in a nonlinear treatment of the SSD, assumes the magnetic field is also a Gaussian random field, but this is not necessary for a kinematic treatment.
}\textsuperscript{,}\footnote{
\Textcite{KopIlySir22, KopKisIly22} have studied the effect of a velocity field whose third cumulant is nonzero.
}
The resulting equations are still quite complicated, and so most analytical work \citep[e.g.][]{kazantsev1968enhancement, ScBoKu02} has additionally assumed that the correlation time of the velocity field is zero (i.e.\@ that it is white noise).\footnote{
\Textcite{vainshtein86} show that if the 4-particle distribution function follows a Fokker-Planck-like equation with diffusion tensor $T_{ij}$, the evolution equation for the longitudinal correlation function of the magnetic field takes a form similar to that in the case of zero correlation time, with the spatial correlation tensor of the velocity field replaced by $T_{ij}$.
However, since they do not give an expression for $T_{ij}$ when the correlation time is nonzero, the effects of a nonzero correlation time are still unclear.
We simply note that such a Fokker-Planck equation can be derived using the methods outlined by \citet{fox1986}.
}

In simulations \citep{BraSub05, kapyla2006strouhal}, the Strouhal number ($\St$, the ratio of the correlation time of the velocity field to its turnover time\footnote{
Note that this definition, which seems to be prevalent in the dynamo community \citep[going back to][eq.~3.14]{KrauseRadler80}, is different from the more common definition which is used for oscillatory flows \citep[e.g.][p.~295]{Whi99}.
}) is typically found to be in the range $0.1 \leq \St \leq 1$.
While this suggests that the effects of a nonzero correlation time are not negligible, it leaves room for hope that perturbative approaches can at least capture the qualitative effects of having a nonzero correlation time.

\citet{bhat2014fluctuation}\footnote{
\Textcite{BhaSub15} describe the same calculation in a more detailed manner.
} and \citet{CarSchSch23} have modelled a velocity field with a nonzero correlation time as a static, smooth flow which is randomly redrawn from an ensemble at fixed intervals of time, say $\tau$ (the `renovating flow' model, first introduced by \citet{ZelMolRuz87}).
They have analytically found that the growth rate is reduced, but the slope of the magnetic energy spectrum
in the Kazantsev range (i.e.\@ the range of wavenumbers much larger than the scale of the velocity field, but much smaller than the resistive scale)
remains unchanged.
The reduction of the growth rate is in qualitative agreement with simulations that use artificial velocity fields \citep{Cha97, MasMalBol11}.
On the other hand, \textcite{kleeorin02}, who also use a renovating flow but do not seem to have performed operator splitting, report that the growth rate is \emph{increased} due to a nonzero correlation time.

While the approach used by \citet{bhat2014fluctuation} and \citet{CarSchSch23} leads to significant computational simplifications, it has a number of shortcomings which limit its generality.
First, a smooth model for the velocity field is applicable only when $\PrM \gg 1$ ($\PrM$, the magnetic Prandtl number, is the ratio of the kinematic viscosity to the magnetic diffusivity).
This is true in some astrophysical contexts (e.g., the interstellar medium), but not in others (e.g., stellar convection zones).
Second, their use of operator splitting is at best justified only when $\Rm$ is large enough that one may neglect $\bigO(\eta \tau)$ terms in the evolution equation for the magnetic correlation function.\footnote{
\Textcite{BhaSub15} claim to show this in their appendix A (note that this appendix is present only in the version of record, not in the preprint).
}
This means that, e.g., their approach cannot be used to understand how the correlation time of the velocity field affects the threshold for onset of the small-scale dynamo.

Assuming the velocity is a Gaussian random field, one can use the Furutsu-Novikov theorem \citep{Fur63, novikov1965} to obtain the evolution equation for the two-point correlation function of the magnetic field as a series in the correlation time (say $\tau_c$) of the velocity field.\footnote{
\Citet{SchekochihinKulsrud2001} discuss how this method is related to other methods such as the cumulant expansion.
}
\citet{SchekochihinKulsrud2001} have used the Furutsu-Novikov theorem to calculate the $\bigO(\tau_c)$ corrections to the growth rates of the single-point moments of the magnetic field.
However, they set the magnetic diffusivity ($\eta$) to zero, rather than taking the $\eta\to 0$ limit; this is known to drastically affect the growth rate of the magnetic field even when $\tau_c = 0$ \parencite[eqs.~1.9, 1.16]{kulsrud92}.
The same problem arises in the work of \textcite{Cha97}, who used a cumulant expansion to calculate the growth rate of the second moment.
Using the Furutsu-Novikov theorem without setting $\eta=0$, we find the $\bigO(\tau_c)$ corrections to the evolution equation for the two-point correlation function of the magnetic field in Fourier space, under the additional assumption that the velocity field is incompressible.

Moving to configuration space, we then obtain the evolution equation for the longitudinal correlation function of the magnetic field when the velocity field is nonhelical (valid for arbitrary $\PrM$).
Assuming a particular form for the longitudinal correlation function of the velocity field (which corresponds to the limit $\PrM \gg 1$) allows us to simplify the evolution equation.
Solving this equation using the WKBJ approximation tells us about the growth rate and the spectral slope of the magnetic field.

In section \ref{section: derivation Fourier}, we derive the evolution equation for the double correlation of the magnetic field in Fourier space.
In section \ref{section: derivation real}, we perform an inverse Fourier transform, and present the evolution equation for the longitudinal correlation function of the magnetic field in nonhelical incompressible turbulence.
In section \ref{section: high Pr}, we simplify the obtained evolution equation by assuming a model for the longitudinal correlation function of the velocity field that is valid at $\PrM \gg 1$.
We then obtain the lowest-order corrections to the growth rate
of the magnetic field and to its spectral slope in the Kazantsev range
due to the correlation time being nonzero.
In section \ref{section: conclusions}, we summarize our results.

The calculations in sections \ref{section: derivation real} and \ref{section: high Pr} were performed using Sympy \citep{sympy2017}.\footnote{
Some enhancements were required, which are available at a fork of the Sympy repository: \url{https://github.com/Kishore96in/sympy/tree/paper_ssdtau_hPr}.
We will attempt to get the required changes (all available on the branch \texttt{paper\_ssdtau\_hPr}) merged into the upstream repository.
}
The scripts and notebooks used for the computations are available on Zenodo \citep{zenodoNotebookSSDtauHPr}.\footnote{
These scripts depend on functions provided by the \texttt{pymfmhd} package (\url{https://github.com/Kishore96in/pymfmhd}).
For convenience, this package is included in the Zenodo upload.
}

\section{Derivation of the evolution equation in Fourier space}
\label{section: derivation Fourier}
\subsection{The induction equation}
Using $\vec{h}(\vec{x}, t)$ to denote the magnetic field and $\vec{w}(\vec{x}, t)$ to denote the velocity field, the induction equation is
\begin{equation}
	\frac{\dh\vec{h}}{\dh t} = \curl\left( \vec{w}\cross\vec{h} \right) + \eta \nabla^2 \vec{h}
\end{equation}
where $\eta$ is the magnetic diffusivity, and boldface denotes a vectorial quantity.

Using a tilde to denote the Fourier transform such that
\begin{equation}
	\FT{f}(\vec{k},t) \defn \int \frac{\d\vec{x}}{(2\pi)^3}\, e^{i\vec{k}\cdot\vec{x}} f(\vec{x},t)
	\label{B2.SSD.shear: FT defn}
\end{equation}
and defining
\begin{align}
	\begin{split}
		\iDelLam_{ijk}^{(\vec{p},\vec{q})}
		\defn{}&
		- i \delta_{ij} p_k
		+ i \delta_{ik} q_j
		\label{B2.SSD.shear: eq: iDelLam defn}
	\end{split}
\end{align}
we write the induction equation as
\begin{align}
	\begin{split}
		\frac{\dh \FT{h}_i^{(\vec{k},t)} }{\dh t}
		={}&
		- \eta k^2 \FT{h}_i^{(\vec{k},t)}
		+ \int_{\substack{ \vec{p}, \vec{q} }} \Dirac^{(\vec{k} - \vec{p} - \vec{q})} \iDelLam_{ijk}^{(\vec{p},\vec{q})} \, \FT{w}_j^{(\vec{p},t)} \, \FT{h}_k^{(\vec{q},t)}
	\end{split} \label{B2.SSD.shear: eq: induction condensed in sheared coord}
\end{align}
where we have assumed the velocity field is incompressible.
Above, and in what follows, we use parenthesized superscripts to denote arguments.
Further, we use the following condensed notation for integrals: $\int_{t', \vec{p}, \vec{q}} \dots \defn \int_{-\infty}^{\infty} \d t' \int\d\vec{p} \int\d\vec{q} \dots$.

We use $\meanBr{\Box}$ to denote the average of a quantity $\Box$.
We assume that the double-correlation of the velocity field is homogeneous and separable, i.e.\@ that it can be written as
\begin{equation}
	\meanBr{ \FT{w}_i^{(\vec{k},t)} \, \FT{w}_j^{(\vec{k'},t')} }
	=
	T_{ij}^{(\vec{k} )} \Dirac^{(\vec{k} + \vec{k'})} \tcf^{(t-t')}
	\label{B2.SSD.shear.tau^1: eq: wiwj homo separable}
	\,,\quad
	2 \int_{0}^\infty \tcf(t) \, \d t = 1
	\,,\quad
	2 \int_{0}^\infty t \, \tcf(t) \, \d t \defn \tau_c
\end{equation}
where $\tau_c$ is the correlation time of the velocity field, and $\tcf(\tau)$ is its temporal correlation function.

\subsection{Evolution equation as a series in \texorpdfstring{$\tau_c$}{tau_c}}
Defining
\begin{equation}
	\mcB_{ij}(\vec{k}, t; \vec{k'}, t')
	\defn
	\FT{h}_i{(\vec{k},t)} \, \FT{h}_j{(\vec{k'},t')}
\end{equation}
we use equation \ref{B2.SSD.shear: eq: induction condensed in sheared coord} to write
\begin{align}
	\begin{split}
		\frac{\dh \mcB_{ij}^{(\vec{k}, t; \vec{k'}, t)} }{\dh t}
		={}&
		- \eta \abs{\vec{k'} }^2 \mcB_{ij}^{(\vec{k}, t; \vec{k'}, t)}
		+ \int_{\substack{ \vec{p}, \vec{q} }} \Dirac^{(\vec{k'} - \vec{p} - \vec{q})} \iDelLam_{jrs}^{(\vec{p},\vec{q})} \, \FT{w}_r^{(\vec{p},t)} \, \mcB_{is}^{(\vec{k}, t; \vec{q}, t)}
		+ \left[ i \exchangeArrow j ;\, \vec{k} \exchangeArrow \vec{k'} \right]
	\end{split} \label{B2.SSD.shear: eq: evolution Bij unaveraged}
\end{align}
where we have used `$\left[ i \exchangeArrow j ;\, \vec{k} \exchangeArrow \vec{k'} \right]$' at the end of the RHS to denote that all the preceding terms should be repeated under the indicated simultaneous relabelling.

We would like to obtain an evolution equation for $\meanBr{\mcB_{ij}}$.
Averaging equation \ref{B2.SSD.shear: eq: evolution Bij unaveraged} gives us
\begin{align}
	\begin{split}
		\frac{\dh \meanBr{ \mcB_{ij}^{(\vec{k}, t; \vec{k'}, t)} } }{\dh t}
		={}&
		- \eta \abs{\vec{k'} }^2 \meanBr{ \mcB_{ij}^{(\vec{k}, t; \vec{k'}, t)} }
		+ \int_{\substack{ \vec{p}, \vec{q} }} \Dirac^{(\vec{k'} - \vec{p} - \vec{q})} \iDelLam_{jrs}^{(\vec{p},\vec{q})} \meanBr{ \FT{w}_r^{(\vec{p},t)} \, \mcB_{is}^{(\vec{k}, t; \vec{q}, t)} }
		+ \left[ i \exchangeArrow j ;\, \vec{k} \exchangeArrow \vec{k'} \right]
	\end{split} \label{eq: evolution Bij averaged}
\end{align}
The evolution equation for $\meanBr{\mcB}$ thus depends on correlations of the form $\meanBr{\FT{w} \mcB}$.
Similarly, the evolution equation for $\meanBr{\FT{w}_{i_1} \dots \FT{w}_{i_n} \mcB}$ would depend on correlations of the form $\meanBr{\FT{w}_{i_1} \dots \FT{w}_{i_n} \FT{w}_{i_{n+1}} \mcB}$, where we have used the shorthand $\FT{w}_{i_\alpha} \defn \FT{w}_{i_\alpha}{(\vec{k}^{(\alpha)},t^{(\alpha)})} $.
Truncating this hierarchy requires additional assumptions (these constitute what is usually referred to as a \emph{closure}).

If we assume $\FT{w}$ is a Gaussian random field, and note that $\mcB$ is a functional of this Gaussian random field (since $\mcB$ at a particular time can depend on $\FT{w}$ at all earlier times), we can use the Furutsu-Novikov theorem (appendix \ref{appendix: FuruNovi}) to simplify the $\meanBr{\FT{w} \mcB}$ terms.
Assuming $\meanBr{\FT{\vec{w}}} = \vec{0}$ and applying the Furutsu-Novikov theorem, we write equation \ref{eq: evolution Bij averaged} as
\begin{align}
	\begin{split}
		\frac{\dh \meanBr{ \mcB_{ij}^{(\vec{k},t; \vec{k'},t)} } }{\dh t}
		={}&
		- \eta \abs{\vec{k'} }^2 \meanBr{ \mcB_{ij}^{(\vec{k},t; \vec{k'},t)} }
		+ \int_{\substack{ \vec{p}, \vec{q}, \vec{k}^{(1)}, t^{(1)} }} \Dirac^{(\vec{k'} - \vec{p} - \vec{q})} \iDelLam_{jrs}^{(\vec{p},\vec{q})} \meanBr{ \FT{w}_r^{(\vec{p},t)} \, \FT{w}_{i_1} } \meanBr{ \frac{\delta \mcB_{is}^{(\vec{k},t; \vec{q},t)} }{\delta \FT{w}_{i_1} } }
		+ \left[ i \exchangeArrow j ;\, \vec{k} \exchangeArrow \vec{k'} \right]
	\end{split} \label{B2.SSD.shear: eq: Bij evolution unsimp condensed}
\end{align}
Our task is now to find an expression for $\meanBr{ \delta \mcB / \delta \FT{w} }$.

To evaluate the $n$-th functional derivative of $\mcB$, we integrate equation \ref{B2.SSD.shear: eq: evolution Bij unaveraged} with respect to time, take $n$ functional derivatives on both sides, average, and use the Furutsu-Novikov theorem.
For notational convenience, we define
\begin{align}
	\begin{split}
		A_{ijk}^{(\vec{k},\vec{k'}; \vec{p}, \vec{q}; t, t')}
		\defn{}&
		e^{ - \eta \left( k^2 + k'^2 \right) \left( t - t' \right) } \Heaviside^{(t-t')} \Dirac^{(\vec{k'} - \vec{p} - \vec{q})} \iDelLam_{ijk}^{(\vec{p}, \vec{q})} 
		\label{B2.SSD.shear: eq: Aijk defn new}
	\end{split}
	\\
	\begin{split}
		\sym_{s; ij; mn}^{(\vec{p'}; \vec{k}, \vec{k'}; \vec{p}, \vec{q}; t, t')}
		\defn{}&
		\delta_{is} \Dirac^{(\vec{k} - \vec{p'})} A_{jmn}^{(\vec{k}, \vec{k'}; \vec{p}, \vec{q}; t, t')}
		+ \delta_{js} \Dirac^{(\vec{k'} - \vec{p'})} A_{imn}^{(\vec{k'}, \vec{k}; \vec{p}, \vec{q}; t, t')}
		\label{B2.SSD.shear: eq: sym defn}
	\end{split}
\end{align}
Recalling that $\meanBr{\vec{w}} = \vec{0}$, the $n$-th functional derivative is given by the following recursion relation:
\begin{align}
	\begin{split}
		\meanBr{ \frac{\delta^n \mcB_{ij}^{(\vec{k}, t; \vec{k'}, t)} }{ \delta \FT{w}_{i_1} \dots \delta \FT{w}_{i_n} } }
		={}&
		\int_{\substack{t', \vec{p}, \vec{q}, \vec{p'}, \\ t^{(n+1)}, \\ \vec{k}^{(n+1)} }} \sym_{s; ij; mn}^{(\vec{p'}; \vec{k}, \vec{k'}; \vec{p}, \vec{q}; t, t')} \meanBr{ \FT{w}_m^{(\vec{p},t')} \FT{w}_{i_{n+1}} } \meanBr{ \frac{\delta^{n+1} \mcB_{sn}^{(\vec{p'}, t'; \vec{q}, t')} }{ \delta \FT{w}_{i_1} \dots \delta \FT{w}_{i_{n+1}} } }
		\\& + \sum_{\alpha=1}^n \int_{\vec{q}, \vec{p'} } \sym_{s; ij; i_\alpha n}^{(\vec{p'}; \vec{k}, \vec{k'}; \vec{k}^{(\alpha)}, \vec{q}; t, t^{(\alpha)} )} \meanBr{ \frac{\delta^{n-1} \mcB_{sn}^{(\vec{p'}, t^{(\alpha)}; \vec{q}, t^{(\alpha)})} }{\delta\FT{w}_{i_1} \dots \delta\FT{w}_{i_{\alpha-1}} \delta\FT{w}_{i_{\alpha+1}} \dots \delta\FT{w}_{i_n} } }
	\end{split} \label{B2.SSD.shear: eq: recursion relation funDer condensed}
\end{align}

Note that equation \ref{B2.SSD.shear: eq: recursion relation funDer condensed} relates the functional derivative of a particular order to other functional derivatives of both higher and lower orders.
Repeated use of equation \ref{B2.SSD.shear: eq: recursion relation funDer condensed} to eliminate all the functional derivatives in equation \ref{B2.SSD.shear: eq: Bij evolution unsimp condensed} thus leads to an infinite series.
Let a particular term in this series contain $m$ time integrals, with the integrand having $n$ factors of the form $\meanBr{ \FT{w} \FT{w} }$ (unequal time correlation) and one factor of the form $\meanBr{\mcB}$.
This term is $\bigO(\tau_c^{m - n})$.
The infinite series we obtain is thus a series in $\tau_c$.
Note that to obtain all the terms at a particular order in $\tau_c$ in this series, one must also use the fact that
\begin{equation}
	\int^t_{-\infty} \d\tau \, f^{(\tau)} \, \tcf^{(t-\tau)}
	=
	f^{(t)} \int^t_{-\infty} \tcf^{(t-\tau)} \, \d\tau
	+ \frac{\d f^{(t)} }{\d t} \int^t_{-\infty} \left( t - \tau \right) \tcf^{(t-\tau)} \, \d\tau 
	+ \bigO(\tau_c^2)
	=
	\frac{1}{2} \, f^{(t)}
	+ \frac{\tau_c}{2} \, \frac{\d f^{(t)} }{\d t}
	+ \bigO(\tau_c^2)
	\label{eq: getting O(tau) terms from Taylor}
\end{equation}
where, for brevity, we have used $f(t)$ to denote the equal-time second-order correlation of the magnetic field.

\subsection{A note on powers of \texorpdfstring{$\tau_c$}{tau_c}}
Consider a simpler model problem, given by (analogous to equation \ref{B2.SSD.shear: eq: Bij evolution unsimp condensed})
\begin{equation}
	\frac{\d X_0}{\d t} =  k E X_1
\end{equation}
where $X_0$ and $X_1$ are the first two variables in a sequence determined by the recursion relation (analogous to equation \ref{B2.SSD.shear: eq: recursion relation funDer condensed})
\begin{equation}
	X_n = \tau E k X_{n+1} + k X_{n-1}
\end{equation}
and $k$, $E$ (analogous to $T_{ij}$), and $\tau$ are constants.
Repeatedly applying the recursion relation to the evolution equation, we find
\begin{align}
	\frac{\d X_0}{\d t} 
	={}&
	k^2 E X_0 + \tau k^2 E^2 X_2
	=
	k^2 E X_0 + \tau k^4 E^2 X_0 + \bigO(\tau^2 k^6 E^3)
\end{align}
Repeated application of the recursion relation has made the RHS a series in $\tau E k^2$.

In the more complicated problem, we use the explicitly appearing factors of $\tau_c$ to keep track of the powers of the actual expansion parameter (let us call it $\tauscl$).
In fact, we abuse notation by using $\bigO(\tau_c^n)$ when we actually mean $\bigO(\tauscl^n)$.

The problem with our abuse of notation becomes evident when one tries to relate $\tauscl$ to the Strouhal number ($\St$): one finds that $\tauscl \propto \St^2$ (appendix \ref{B2.SSD.real: appendix: Rm and St in terms of tau and eta}).
This is because $T_{ij}(0) \scalesAs \tau_c u_\text{rms}^2$ (the factor of $\tau_c$ comes from $\tcf$ in equation \ref{B2.SSD.shear.tau^1: eq: wiwj homo separable}).
Despite this problem, we use this notation to allow easy comparison of our work with previous work \citep{SchekochihinKulsrud2001, bhat2014fluctuation}.

\subsection{Evolution equation with small (nonzero) correlation time}
Repeatedly using equation \ref{B2.SSD.shear: eq: recursion relation funDer condensed} to eliminate all the functional derivatives in equation \ref{B2.SSD.shear: eq: Bij evolution unsimp condensed} and neglecting $\bigO(\tau_c^2)$ terms, we obtain an extremely long evolution equation, given in appendix \ref{B2.SSD.shear: eq: Bij evolution in terms of I* new}.
In this evolution equation, the dependence on the temporal correlation function of the velocity field only enters through the constants $g_1$ and $g_2$, defined as
\begin{subequations}
\begin{align}
	g_1 \defn{}& \frac{1}{\tau_c} \int_{-\infty}^t \d t' \int_{-\infty}^{ t' } \d t_1 \int_{-\infty}^{ t_1 } \d t_2 \, \tcf^{(t-t_1)} \tcf^{(t'-t_2)}
	\\
	g_2 \defn{}& \frac{1}{\tau_c} \int_{-\infty}^t \d t' \int_{-\infty}^{ t' } \d t_2 \int_{-\infty}^{ t_2 } \d t_1 \, \tcf^{(t-t_1)} \tcf^{(t'-t_2)}
\end{align} \label{B2.SSD.shear.tau^1: eq: g1 g2 defn}
\end{subequations}
In appendix \ref{appendix: g_1 + g_2}, we show that $g_1 + g_2 = 1/4$ regardless of the form of $\tcf(t)$.
Table \ref{table: g1 g2 values} gives $g_2$ for some forms of the temporal correlation function.

\begin{table}
	\centering
	\begingroup
		\renewcommand{\arraystretch}{2}
		\begin{tabular}{llll}
			Name & $\tcf(\tau)$ &  $g_2$ \\
			\hline
			Exponential & $\displaystyle \frac{1}{2\tau_c} \, e^{-\abs{\tau}/\tau_c} $ & 1/8 \\
			Top hat & $\displaystyle \frac{1}{4\tau_c} \, \Heaviside{\!\left( 2\tau_c - \tau \right)} \Heaviside{\!\left( \tau + 2 \tau_c \right)}$ & 1/12
		\end{tabular}
	\endgroup
	\caption{
		Values of $g_2$ for some temporal correlation functions.
		}
	\label{table: g1 g2 values}
\end{table}

\section{The evolution equation in real space}
\label{section: derivation real}
\subsection{Definition and properties of the longitudinal correlation function}
When the magnetic field is homogeneous, isotropic, and mirror-symmetric, the double-correlation of the magnetic field in real space ($M_{ij}(\vec{r}) \defn \meanBr{h_i(\vec{r},t) \, h_j(\vec{0},t) }$) can be written as\footnote{
See \citet[eq.~5.63]{lesieur2008turbulence} and \citet[eq.~23]{vainshtein86} for a general expression that does not assume mirror-symmetry.
}
\begin{equation}
	M_{ij}(\vec{r})
	=
	\left( \delta_{ij} - \frac{r_i r_j}{r^2} \right) M_N(r)
	+ \frac{r_i r_j}{r^2} \, M_L(r)
\end{equation}
with
\begin{equation}
	M_N = \frac{1}{2r}\,\frac{\dh}{\dh r}{\left( r^2 M_L\right)}
\end{equation}
where $M_L$ is the \emph{longitudinal correlation function}.

On the other hand, in Fourier space, one can write the double-correlation of a homogeneous, isotropic, and mirror-symmetric magnetic field as \citep[see][eq.~3.4.12]{batchelor1953homoturb}
\begin{equation}
	M_{ij}(\vec{k})
	=
	\op{P}_{ij}(\vec{k}) \, M(k)
	\label{B2.SSD.real: eq: Mij fourier}
	\,,\quad
	\op{P}_{ij}(\vec{k}) \defn \delta_{ij} - \frac{k_i k_j}{k^2}
\end{equation}
We find that
\begin{equation}
	2 M(r) = M_{ii}(r) = \frac{1}{r^2} \, \frac{\dh}{\dh r}{\left( r^3 M_L \right)}
	\label{B2.SSD.real: eq: M M_L relation}
\end{equation}
where $M(r)$ denotes the 3D inverse Fourier transform of $M(k)$, which is explicitly given by \citep[eq~12.4]{MoninYaglomVol2}
\begin{equation}
	M(r)
	=
	\frac{4\pi}{r} \int_0^\infty \d k \, M(k) \, k \sin(kr)
	\label{B2.SSD.real: eq: spherically symmetric IFT}
\end{equation}
Inverting equation \ref{B2.SSD.real: eq: M M_L relation}, we have (appendix \ref{appendix: M_L M relation lower limit} discusses the value of the lower limit on the RHS)
\begin{equation}
	M_L = \frac{1}{r^3} \int_0^r r^2 M_{ii}(r) \, \d r
	\label{B2.SSD.real: eq: M_L M relation}
\end{equation}

In what follows, for the velocity correlation (see equation \ref{B2.SSD.shear.tau^1: eq: wiwj homo separable}), we also use an expression that follows from homogeneity, isotropy, mirror-symmetry,\footnote{
Note that kinetic helicity can significantly affect the growth rate of the small-scale dynamo when $\PrM$ is not large \citep{MalBol10}.
} and incompressibility \citep[eq.~3.4.12]{batchelor1953homoturb}:
\begin{equation}
	T_{ij}(\vec{k})
	=
	\op{P}_{ij}(\vec{k}) \, E(k)
\end{equation}
Similar to the case of the magnetic field, we define the longitudinal correlation function of the velocity field, $E_L(r)$, by
\begin{equation}
	E_L = \frac{1}{r^3} \int_0^r r^2 T_{ii}(r) \, \d r
	\label{B2.SSD.real: eq: E_L T relation}
\end{equation}
The inverse of this relation is
\begin{equation}
	T_{ii}(r) = \frac{1}{r^2} \, \frac{\dh}{\dh r}{\left( r^3 E_L \right)}
	\label{B2.SSD.real: eq: T E_L relation}
\end{equation}

Assuming $\left[ r \, \d E_L/\d r \right]_{r=0} = 0$, equation \ref{B2.SSD.real: eq: T E_L relation} implies
\begin{equation}
	\meanBr{ w_i{(\vec{x}, t)} \, w_i{(\vec{x}, t)} }
	=
	3 \, E_L(0) \, \tcf(0)
	\label{eq: urms EL tcf relation}
\end{equation}

\subsection{Evolution equation when the velocity field is nonhelical}
Using the identities in appendix \ref{appendix: fourier identities}, we take the inverse Fourier transform of equation \ref{B2.SSD.shear: eq: Bij evolution in terms of I* new} and contract $j$ with $i$.
Using equations \ref{B2.SSD.real: eq: M M_L relation}, \ref{B2.SSD.real: eq: M_L M relation}, \ref{B2.SSD.real: eq: E_L T relation}, and \ref{B2.SSD.real: eq: T E_L relation}, we then obtain an evolution equation for $M_L(r,t)$ (equation \ref{B2.SSD.real: eq: dMLdt nonhelical} in appendix \ref{appendix: evol eq g nonzero}).
This equation contains the third and fourth spatial derivatives of $M_L(r,t)$.

In appendix \ref{appendix: g_1 + g_2}, we prove that the constants, $g_1$ and $g_2$, which depend on the form of $\tcf(t)$, are related as $g_1 + g_2 = 1/4$, regardless of the form of $\tcf(t)$.
Accounting for this, the coefficient of $\dh^4 M_L/\dh r^4$ becomes zero, and we obtain the following evolution equation for the longitudinal correlation function of the magnetic field:
\begin{align}
	\begin{split}
		\frac{\dh M_L}{\dh t}
		={}&
		\frac{1}{r^4} \, \frac{\dh}{\dh r}{\left( \left[ \kappa(r) + \tau_c \, \kappa_\tau(r) \right] r^4 \, \frac{\dh M_L}{\dh r} \right)}
		+ \left[ G(r) + \tau_c \, G_\tau(r) \right] M_L
		+ \tau_c \eta \left[
			- \frac{4 }{r^{5}} \, \frac{\d}{\d r}{\left( r^{4} \, \frac{\d S_{2} }{\d r} \right)} \, \frac{\dh M_L}{\dh r}
			+ \frac{\d S_2}{\d r} \, \frac{\dh^3 M_L}{\dh r^3}
			\right]
	\end{split} \label{B2.SSD.real: eq: dMLdt nonhelical with g=0}
\end{align}
where we have defined
\begin{subequations}
\begin{align}
	\begin{split}
		S_2(r)
		\defn{}&
		2 \left( E_L(0) - E_L(r) \right)
	\end{split}
	\\
	\begin{split}
		\kappa(r)
		\defn{}&
		2 \eta + E_L(0) - E_L(r)
	\end{split}
	\\
	\begin{split}
		G(r)
		\defn{}&
		- \frac{\d^2 E_L}{\d r^2} - \frac{4}{r} \, \frac{\d E_L}{\d r}
	\end{split}
	\\
	\begin{split}
		\kappa_\tau(r)
		={}&
		g_2 \left[
			- 8 v_{2} E_{L}{\left(0 \right)} 
			+ \left( E_{L}'(r) \right)^{2} 
			- \frac{1}{r^{4}} \, \frac{\d}{\d r}{\left( r^{4} \frac{\d{\left( E_{L}^{2} \right)} }{\d r} \right)}
			\right]
		\\& - 4 \eta v_{2}
		+ \frac{3 \eta }{2 r^{4}} \, \frac{\d}{\d r}{\left( r^{4} \, \frac{\d S_{2} }{\d r} \right)}
		+ \frac{\left( S_{2}'(r) \right)^{2}}{16}
		- \frac{1}{16 r^{4}} \, \frac{\d}{\d r}{\left( r^{4} \, \frac{\d \left( S_{2}^{2} \right) }{\d r} \right)}
	\end{split}
	\\
	\begin{split}
		G_\tau(r)
		={}&
		g_2 \Bigg[ 
			- 2 E_{L}{\left(r \right)} E_{L}''''{\left(r \right)} 
			- 6 E_{L}'{\left(r \right)} E_{L}'''{\left(r \right)} 
			- 4 \left( E_{L}''{\left(r \right)}\right)^{2} 
			- \frac{16 E_{L}{\left(r \right)} E_{L}'''{\left(r \right)}}{r} 
			\\&\continuedTerm - \frac{40 E_{L}'{\left(r \right)} E_{L}''{\left(r \right)}}{r} 
			- \frac{16 E_{L}{\left(r \right)} E_{L}''{\left(r \right)}}{r^{2}} 
			- \frac{16 \left(E_{L}'{\left(r \right)}\right)^{2}}{r^{2}} 
			+ \frac{16 E_{L}{\left(r \right)} E_{L}'{\left(r \right)}}{r^{3}}
			\Bigg]
		\\& + \frac{ \eta \, S_{2}''''{\left(r \right)} }{2}
		+ \frac{4 \eta \, S_{2}'''{\left(r \right)}}{r}
		+ \frac{4 \eta \, S_{2}''{\left(r \right)}}{r^{2}}
		- \frac{4 \eta \, S_{2}'{\left(r \right)}}{r^{3}}
		- \frac{S_{2}{\left(r \right)} S_{2}''''{\left(r \right)} }{8}
		- \frac{3 S_{2}'{\left(r \right)} S_{2}'''{\left(r \right)} }{8}
		\\& - \frac{\left( S_{2}''{\left(r \right)}\right)^{2} }{4}
		- \frac{S_{2}{\left(r \right)} S_{2}'''{\left(r \right)} }{r} 
		- \frac{5 S_{2}'{\left(r \right)} S_{2}''{\left(r \right)}}{2 r}
		- \frac{ S_{2}{\left(r \right)} S_{2}''{\left(r \right)}}{r^{2}}
		- \frac{ \left( S_{2}'{\left(r \right)}\right)^{2}}{r^{2}}
		+ \frac{ S_{2}{\left(r \right)} S_{2}'{\left(r \right)}}{r^{3}}
	\end{split}
\end{align} \label{eq: coefficients in evolution equation g=0}%
\end{subequations}
Above, $v_2$ \citep[also defined by][eq.~9]{kazantsev1968enhancement} is given by
\begin{equation}
	v_2
	=
	- \frac{1}{12} \left[ \nabla^{2}{\left( \frac{1}{r^2} \, \frac{\d}{\d r}{\left( r^3 E_L \right)} \right)} \right]_{r=0}
	\label{B2.SSD.real: eq: vi in terms of real-space}
\end{equation}
Note that $g_2$ is the only surviving parameter that depends on the form of the temporal correlation function.
Also note that as anticipated by \textcite{vainshtein86}, the non-resistive $\bigO(\tau_c)$ corrections do not change the form of the evolution equation for $M_L$.

In general, one obtains an additional term
\begin{equation}
	- \frac{8}{r^3} \left( \eta + 2 g_2 \, E_L(0) \right) M_L(0) \left[ \frac{\d E_L}{\d r} \right]_{r = 0}
	\label{B2.SSD.real: eq: dMLdt extra ML(0) term}
\end{equation}
on the RHS of equation \ref{B2.SSD.real: eq: dMLdt nonhelical with g=0}.
$E_L(r)$ is usually expected to have zero slope at the origin, and so we ignore this term.

\subsection{Comparison with previous results}
\label{B2.SSD.real: compare tau=0 with previous results}
Our evolution equation for the longitudinal correlation function (equation \ref{B2.SSD.real: eq: dMLdt nonhelical with g=0}) agrees with that derived by \citet[eq.~56]{ScBoKu02} on setting $\tau_c=0$.
Accounting for the fact that \citet[eq.~10]{vainshtein86} and \citet{Sub97} use a slightly different definition of the longitudinal correlation function of the velocity field (such that their $T_{LL} = E_L/2$), we also find that our equations are consistent with theirs.\footnote{
\Citet{Sub97} points out a sign error in the equations given by \citet{vainshtein86}.
}

\Textcite[eq.~17]{bhat2014fluctuation} have derived an evolution equation for the longitudinal correlation of the magnetic field in a homogeneous, isotropic, and nonhelical renovating flow (explicitly neglecting $\bigO(\eta\tau_c)$ terms which we have retained).
They seem to use the Taylor expansion $f(\tau) = f(0) + \tau \d f/\d t + \bigO(\tau^2)$ to evaluate the time derivative of a correlation function of the magnetic field (denoted here as $f$), given an expression for $f(\tau)$ in terms of $f(0)$ (see the discussions above equations 3.11 and 3.21 of \citealp{BhaSub15}).
However, the error in such an estimate of $\d f/\d t$ is proportional to $\tau \d^2 f/\d t^2$, i.e.\@ their final evolution equation misses some $\bigO(\tau)$ terms.
The effect of this seems similar to neglecting the $\bigO(\tau_c)$ term in our equation \ref{eq: getting O(tau) terms from Taylor}.
This term is responsible for all the $\bigO(\tau_c)$ terms in appendix \ref{appendix: ssdShTerms} that are independent of $g_1$ and $g_2$.
Dropping these terms corresponds to setting $g = g_1 + g_2$ (which is nonzero) in equation \ref{B2.SSD.real: eq: dMLdt nonhelical} (appendix \ref{appendix: evol eq g nonzero}).
This suggests the reason for $\dh^4 M_L / \dh r^4$ appearing with a nonzero $\bigO(\tau_c)$ coefficient in their final evolution equation, contrary to our result (equation \ref{B2.SSD.real: eq: dMLdt nonhelical with g=0}).
Additionally, they found no $\tau_c$-dependent corrections to the coefficient of $M_L$, while we do.

\Textcite[eq.~5]{kleeorin02} have also derived such an equation using a renovating flow, but without operator splitting (instead, they assume that the velocity field is Gaussian random).\footnote{
Note that \textcite[p.~2]{bhat2014fluctuation} suggest that \textcite[cf.~eq.~B1]{kleeorin02} wrongly dropped some terms in a Taylor expansion.
\label{footnote: Kleeorin mistake}
}
Just like us, they obtain $\tau_c$-dependent corrections to the coefficient of $M_L$, and moreover do not obtain any terms dependent on $\dh^4 M_L/\dh r^4$.
However, their coefficient of $\dh^3 M_L / \dh r^3$ seems to be independent of $\eta$, unlike in our case where the coefficient is $\bigO(\tau_c \eta)$.

\section{Simplification at high Prandtl number}
\label{section: high Pr}
\subsection{Simplified evolution equation}
We model $E_L(r) = E_0 \exp(- k_f^2 r^2/2)$.\footnote{
Since equation \ref{B2.SSD.real: eq: dMLdt nonhelical with g=0} contains fourth-order derivatives of $E_L$, this is different from simply taking $E_L(r) = E_L(0) \left( 1 - k_f^2 r^2/2\right)$.
}
We change the temporal and spatial variables and define the following quantities:
\begin{equation}
		T \defn E_0 k_f^2 t
	\,,\quad
		R \defn k_f r
	\,,\quad
		\etascl \defn \frac{\eta}{E_0}
	\,,\quad
		\tauscl \defn \tau_c k_f^2 E_0
	\,,\quad
		\widetilde{S}_2(R)
		\defn
		\frac{S_2(r)}{E_0}
	\,,\quad
		\widetilde{\kappa}(R)
		\defn
		\frac{\kappa(r)}{ E_0 }
\end{equation}
\begin{equation}
		\widetilde{G}(R)
		\defn
		\frac{G(r) }{k_f^2 E_0 }
	\,,\quad
		\widetilde{\kappa}_\tau(R)
		\defn
		\frac{\kappa_\tau(r) }{ k_f^2 E_0^2 }
	\,,\quad
		\widetilde{G}_\tau(R)
		\defn
		\frac{ G_\tau(r) }{ k_f^4 E_0^2}
\end{equation}

To simplify the equation, we further assume that $\Rm \gg 1$, so that the magnetic field grows the fastest on scales much smaller than the integral scale (i.e.\@ $R \ll 1$).
We thus expand all the coefficients as series in $R$ and discard $\bigO(R M_L)$ terms (where $\dh M_L/\dh R \scalesAs \bigO(R^{-1} M_L)$).
The evolution equation for $M_L$ (equation \ref{B2.SSD.real: eq: dMLdt nonhelical with g=0}) then becomes
\begin{align}
	\begin{split}
		\frac{\dh M_L}{\dh T}
		={}&
		\frac{1}{R^4} \, \frac{\dh}{\dh R}{\left( \left[ \widetilde{\kappa}(R) + \tauscl \widetilde{\kappa}_\tau(R) \right] R^4 \, \frac{\dh M_L}{\dh R} \right)}
		+ \left[ \widetilde{G}(R) + \tauscl \widetilde{G}_\tau(R) \right] M_L
		\\& + \tauscl \left( 28 \etascl R - \frac{40 \etascl}{R} \right) \frac{\dh M_L}{\dh R}
		+ \tauscl \etascl \left( 2 R - R^{3} \right) \frac{\dh^3 M_L}{\dh R^3}
	\end{split} \label{B2.SSD.real: eq: dMLdt nonhelical highPr}
\end{align}
with
\begin{align}
	\begin{split}
		\widetilde{\kappa}(R)
		={}&
		\frac{R^{2}}{2} + 2 \etascl
	\end{split}
	\\
	\begin{split}
		\widetilde{G}(R)
		={}&
		5
	\end{split}
	\\
	\begin{split}
		\widetilde{\kappa}_\tau(R)
		={}&
		10 \etascl - R^{2} \left( \frac{21 \etascl}{2} + 13 g_{2} + \frac{3}{2}\right)
	\end{split}
	\\
	\begin{split}
		\widetilde{G}_\tau(R)
		={}&
		- 35 \etascl - 130 g_{2} - 15
	\end{split}
\end{align}

\subsection{WKBJ analysis}
\subsubsection{Elimination of higher derivatives}
To apply the WKB method, we need to express the derivatives of order greater than two in equation \ref{B2.SSD.real: eq: dMLdt nonhelical highPr} in terms of lower derivatives.
Since these higher derivatives only appear multiplied by $\tauscl$, one can use the `Landau-Lifshitz' approach \citetext{\citealp[sec.~75]{LandauLifshitzVol2}; \citealp[p.~4]{bhat2014fluctuation}} and eliminate them perturbatively as follows.
Assuming $M_L(r,t) = \widetilde{M}_L(R) \exp(\gamma T)$, setting $\tauscl = 0$, and taking derivatives wrt.\@ $R$ of equation \ref{B2.SSD.real: eq: dMLdt nonhelical highPr}, we obtain an expression for the third derivative of $\widetilde{M}_L$ in terms of the lower-order derivatives.
Substituting this expression in equation \ref{B2.SSD.real: eq: dMLdt nonhelical highPr} and discarding $\bigO(\etascl^2)$ terms, we obtain
\begin{multline}
	\left[ 5 - \gamma + \tauscl \left(- 35 \etascl - 130 g_{2} - 15\right) \right] \widetilde{M}_L(R)
	+ \left[ \frac{8 \etascl}{R} + 3 R + \tauscl \left(\frac{\etascl \left(4 \gamma - 32\right)}{R} + R \left(\etascl \left(- 2 \gamma - 19\right) - 78 g_{2} - 9\right)\right) \right] \frac{\d \widetilde{M}_L}{\d R}
	\\ + \left[ 2 \etascl + \frac{R^{2}}{2} + \tauscl \left(- 6 \etascl + R^{2} \left(- \frac{5 \etascl}{2} - 13 g_{2} - \frac{3}{2}\right)\right) \right]  \frac{\d^2 \widetilde{M}_L}{\d R^2}
	= 0
	\label{B2.SSD.real.nonhel: eq: dMLdt WKB without higher derivatives}
\end{multline}

\subsubsection{Change of variables}
We change the variable of differentiation to
\begin{equation}
	x \defn \log{\!\left( R /\sqrt{\etascl} \right)}
	\label{B2.SSD.real.nonhel.hPr.WKB: eq: x defn}
\end{equation}
with $\Upsilon(x) \defn \widetilde{M}_L(R)$ and obtain
\begin{equation}
	A_2(x) \, \frac{\d^2 \Upsilon}{\d x^2} + A_1(x) \, \frac{\d \Upsilon}{\d x} + A_0(x) \, \Upsilon(x) = 0
	\label{B2.SSD.nonhel: eq: Upsilon 2nd order diffeq}
\end{equation}
with
\begin{subequations}
\begin{align}
	\begin{split}
		A_0(x)
		\defn{}&
		5 - \gamma + \tauscl \left(- 35 \etascl - 130 g_{2} - 15\right)
	\end{split}
	\\
	\begin{split}
		A_1(x)
		\defn{}&
		\frac{5}{2} + 6 e^{- 2 x}
		+ \tauscl \left( - \etascl \left[ 2 \gamma + \frac{33}{2}\right] - 65 g_{2} + 4 \gamma e^{- 2 x} - \frac{15}{2} - 26 e^{- 2 x}\right)
		\label{B2.SSD.real.nonhel.hPr.WKB: eq: A1}
	\end{split}
	\\
	\begin{split}
		A_2(x)
		\defn{}&
		\frac{1}{2} + 2 e^{- 2 x}
		+ \tauscl \left(- \frac{5 \etascl}{2} - 13 g_{2} - \frac{3}{2} - 6 e^{- 2 x}\right)
		\label{B2.SSD.real.nonhel.hPr.WKB: eq: A2}
	\end{split}
\end{align}
\label{B2.SSD.real.nonhel.hPr.WKB: eq: A0,1,2}
\end{subequations}

\subsubsection{Conversion to Schr\"odinger-like form}
Further substituting $\Upsilon(x) = \beta(x) \, \Theta(x)$, we find that imposing
\begin{equation}
	\frac{\d\beta}{\d x} = - \beta(x) \, \frac{A_1(x)}{2 A_2(x)}
	\label{B2.SSD.real.nonhel: eq: beta diffeq}
\end{equation}
gives us
\begin{equation}
	\frac{\d^2\Theta}{\d x^2} + p(x) \, \Theta(x) = 0
	\label{B2.SSD.real.nonhel.hPr.WKB: eq: WKB form}
\end{equation}
with
\begin{align}
	\begin{split}
		p(x)
		\defn{}&
		\frac{A_{0}{\left(x \right)}}{A_{2}{\left(x \right)}} - \frac{A_{1}^{2}{\left(x \right)}}{4 A_{2}^{2}{\left(x \right)}} + \frac{A_{1}{\left(x \right)} }{2 A_{2}^{2}{\left(x \right)}} \, \frac{\d A_{2} }{\d x} - \frac{1}{2 A_{2}{\left(x \right)}} \, \frac{\d A_{1} }{\d x}
	\end{split} \label{B2.SSD.real.nonhel.hPr.WKB: eq: p general}
\end{align}
The WKB solutions for $\Theta(x)$ are then
\begin{equation}
	\Theta(x) \propto \abs{ p(x) }^{-1/4} \exp{\!\left( \pm i \int^x \sqrt{ p(x) } \, \d x\right)}
	\label{B2.SSD.real.nonhel.hPr.WKB: eq: Theta general}
\end{equation}
Appendix \ref{section: WKB validity} discusses the validity of the WKB approximation for this problem.

\subsubsection{Asymptotic solution at \texorpdfstring{$x\to -\infty$}{x -> -oo}}
We first note that
\begin{equation}
	\lim_{x\to-\infty} \frac{A_1(x)}{A_2(x)} = 3 + \bigO(\tau_c)
\end{equation}
Equation \ref{B2.SSD.real.nonhel: eq: beta diffeq} then implies
\begin{equation}
	\beta(x) \propto e^{- 3x/2 + \bigO(\tau_c) } \,,\quad x\to-\infty
\end{equation}
Physically, we expect $\widetilde{M}_L(r) \defn \beta(x) \, \Theta(x)$ to approach a constant value as $r\to 0$ ($x\to-\infty$).
Since
\begin{equation}
	\lim_{x\to-\infty} p(x) = - \frac{9}{4}  + \bigO(\tau_c)
\end{equation}
we need to pick the $\exp(- i \int\dots)$ branch, which gives us
\begin{equation}
	\Theta(x) \propto e^{3x/2 + \bigO(\tau_c) } \,, \quad x\to-\infty
\end{equation}

\subsubsection{Asymptotic solution at \texorpdfstring{$x\to \infty$}{x -> oo}}
On the other hand, at $x\to\infty$, equation \ref{B2.SSD.real: eq: dMLdt nonhelical highPr} (which we used to derive equation \ref{B2.SSD.real.nonhel: eq: dMLdt WKB without higher derivatives}) becomes invalid, and so we have to go back to equation \ref{B2.SSD.real: eq: dMLdt nonhelical with g=0}.
Replacing $E_L(r\ne 0) = 0$, setting $\tauscl = 0$, and taking $M_L(r,t) = \widetilde{M}_L(R) \exp{\!\left( \gamma T \right)}$, equation \ref{B2.SSD.real: eq: dMLdt nonhelical with g=0} reduces to
\begin{equation}
	\left( 1 + 2 \etascl \right) \frac{\d^2 \widetilde{M}_L}{\d R^2} + \frac{4 \left( 1 + 2\etascl \right) }{R} \, \frac{\d \widetilde{M}_L}{\d R} - \gamma \widetilde{M}_L(R) = 0
\end{equation}
$p(x)$ (equation \ref{B2.SSD.real.nonhel.hPr.WKB: eq: p general}) is then given by
\begin{equation}
	p(x)
	=
	- \frac{\etascl \gamma }{2 \etascl + 1} \, e^{2 x}
	+ \bigO(1, x\to \infty)
\end{equation}

We have (setting $\tau_c= 0$)
\begin{equation}
	\lim_{x \to \infty} \frac{A_1(x)}{A_2(x)} = 3
\end{equation}
This implies
\begin{equation}
	\beta(x) \propto e^{-3x/2} \,,\quad x \to \infty
\end{equation}
Recall that equation \ref{B2.SSD.real: eq: dMLdt nonhelical highPr} (the evolution equation which we are analyzing) is valid when $\Rm \gg 1$.
It is thus reasonable to assume $\gamma > 0$ (i.e.\@ that there exists a growing solution).
Since we require $\widetilde{M}_L(r) \defn \beta(x) \, \Theta(x)$ to approach zero as $r \to \infty$ ($x \to \infty$), we need to pick the $\exp(+ i \int\dots)$ branch, which gives us
\begin{equation}
	\Theta(x) \propto \exp{\!\left( - e^{x} \sqrt{ \frac{\etascl \gamma }{2 \etascl + 1} } \right)}
	\,,\quad x \to \infty
\end{equation}

\subsubsection{Connection formulae}
Above, we saw that different solution branches need to be chosen at $\pm\infty$ in order to satisfy the boundary conditions.
This means we must have a turning point.
Since $\lim_{x\to - \infty} p(x) < 0$ and $\lim_{x\to \infty} p(x) < 0$, there must be at least two turning points (say $x_1$ and $x_2$), between which $p(x) > 0$.

Let us write the general solution as
\begin{equation}
	\Theta(x)
	=
	\begin{dcases}
		\frac{ C_1 }{ \abs{ p }^{1/4} } \, e^{ - i \int_{x_1}^x \sqrt{p} \, \d x } & x < x_1
		\\
		\frac{ C_{+} }{ \abs{ p }^{1/4} } \, e^{ i \int_{x_1}^x \sqrt{p} \, \d x }
		+ \frac{ C_{-} }{ \abs{ p }^{1/4} } \, e^{ - i \int_{x_1}^x \sqrt{p} \, \d x }
		& x_1 < x < x_2
		\\
		\frac{ C_2 }{ \abs{ p }^{1/4} } \, e^{ i \int_{x_2}^x \sqrt{p} \, \d x } & x_2 < x
	\end{dcases}
\end{equation}
and number the regions above as I, II, and III respectively.
Going from II to I, we find
\begin{equation}
	C_1 = C_{+} \, e^{-i\pi/4} = C_{-} \, e^{i\pi/4}
\end{equation}
We thus write
\begin{equation}
	\Theta(x)
	=
	\frac{ C_1 \sqrt{2} }{ \abs{ p }^{1/4} } \cos{\!\left( \int_{x_1}^x \sqrt{p} \, \d x \right)}
	\,,\quad x_1 < x < x_2
	\label{B2.SSD.real.nonhel.hPr.WKB: eq: Theta region II from connection formulae}
\end{equation}
Now, considering this near $x_2$, defining $P \defn \int_{x_1}^{x_2} \sqrt{p} \, \d x$, and going from II to III, we find
\begin{equation}
	C_2
	= C_{+} \, e^{iP + i\pi/4}
	= C_{-} \, e^{-iP -i\pi/4}
\end{equation}
Recalling that $C_1 = C_{+} \, e^{-i\pi/4}$, we write the first equality above as
\begin{equation}
	C_2 = C_1 \, e^{iP + i\pi/2}
\end{equation}
By requiring that $C_1$ and $C_2$ be real, we find that we need
\begin{equation}
	\int_{x_1}^{x_2} \sqrt{p(x)} \, \d x = \frac{\left(2n + 1\right) \pi}{2}
	\label{B2.SSD.real.nonhel.hPr.WKB: eigencondition}
\end{equation}
where $n$ is any nonnegative integer.

\subsubsection{Estimate of the growth rate}
If we are interested in scales much above the resistive scale (but still below the integral scale, since we have already assumed $R \ll 1$), we can assume $e^{-2x} \ll 1$.
We thus expand $p(x)$ (equation \ref{B2.SSD.real.nonhel.hPr.WKB: eq: p general}) about $x=\infty$ and neglect $\bigO(e^{-2x}, x\to\infty)$ terms.
In what follows, we define
\begin{equation}
	\Delta \defn x_2 - x_1
	= \log(R_2/R_1)
	> 0
\end{equation}
where $R_1$ and $R_2$ are the values of $R$ corresponding to $x_1$ and $x_2$.
The square of the integral on the RHS of equation \ref{B2.SSD.real.nonhel.hPr.WKB: eigencondition} can then be estimated as\footnote{
While estimating the integral, it is convenient to change the variable of integration to $y\defn \exp(-2x)$ (such that $y = \etascl/R^2$).
}
\begin{equation}
	\left( \int_{x_1}^{x_2} \sqrt{p(x)} \, \d x \right)^2
	=
	\frac{\Delta^{2} \left( 15 - 8 \gamma + \tauscl \gamma \left(- 208 g_{2} - 24 \right) \right)}{4}
	\label{B2.SSD.real.nonhel.hPr.WKB: eq: sq int sqrt p estimate}
\end{equation}
Squaring both sides of equation \ref{B2.SSD.real.nonhel.hPr.WKB: eigencondition}, choosing $n=0$,\footnote{
The value of $n$ only affects the $\bigO(\Delta^{-2})$ corrections to the growth rate.
} and iteratively solving for $\gamma$, we find\footnote{
The growth rate is reduced when $\tauscl \ne 0$ if $g_2 > -3/26$.
We have not been able to find any general proof that $g_2$ satisfies this inequality.
While $g_2 > 0$ if $\tcf(t)$ is nonnegative, note that $\tcf(t)$ is allowed to be negative for some $t \ne 0$.
}
\begin{equation}
	\gamma = 
	\frac{15}{8} - \tauscl \left( \frac{195 g_{2}}{4} + \frac{45}{8}\right)
	+ \bigO{\left( \Delta^{-2} \right)}
	+ \bigO(\tauscl^2)
	\label{B2.SSD.real.nonhel.hPr.WKB: eq: gamma soln}
\end{equation}

\subsubsection{WKB solution for the correlation function}
As in the estimation of the growth rate, we assume $e^{-2x} \ll 1$ and thus neglect $\bigO(e^{-2x}, x\to \infty)$ terms below.
This allows us to write
\begin{align}
	\begin{split}
		\int_{x_1}^x \sqrt{p(x)} \d x
		={}&
		\frac{\pi \left( x - x_1 \right) }{ 2\Delta } + \bigO(\tauscl^2)
		+ \bigO(e^{-2x}, x\to \infty)
	\end{split}
	\\
	\begin{split}
		p(x)
		={}&
		\frac{\pi^2}{ 4 \Delta^2 }
		+ \bigO(\tauscl^2)
		+ \bigO(e^{-2x}, x\to \infty)
	\end{split}
\end{align}
Equation \ref{B2.SSD.real.nonhel.hPr.WKB: eq: Theta region II from connection formulae} then implies that for $x_1 < x < x_2$, we have
\begin{equation}
	\Theta(x)
	\propto
	\cos{\!\left( \frac{\pi}{2\Delta} \log(R/R_1) + \bigO(\tauscl^2) \right)}
\end{equation}

Using equation \ref{B2.SSD.real.nonhel.hPr.WKB: eq: gamma soln} for $\gamma$ and recalling the definitions of $A_1$ (equation \ref{B2.SSD.real.nonhel.hPr.WKB: eq: A1}) and $A_2$ (equation \ref{B2.SSD.real.nonhel.hPr.WKB: eq: A2}), we write
\begin{equation}
	\frac{ A_1(x) }{ A_2(x) }
	=
	5 + \etascl \tauscl \left( \frac{\pi^{2} }{ 2 \Delta^{2} } - \frac{31}{2} \right)
	+ \bigO(\tauscl^2)
\end{equation}
Equation \ref{B2.SSD.real.nonhel: eq: beta diffeq} then tells us that
\begin{equation}
	\beta(x)
	=
	\exp{\!\left\{ - \left[ 5/2 + \bigO(\etascl\tauscl) \right] x \right\}}
	\propto
	R^{-5/2 + \bigO(\etascl\tauscl)}
\end{equation}
Note that $\etascl \tauscl \propto \St/\Rm$ (appendix \ref{B2.SSD.real: appendix: Rm and St in terms of tau and eta}).
Recalling that $M_L(r) = \beta(x) \, \Theta(x) \, e^{\gamma T}$, we write
\begin{equation}
	M_L(r, t)
	=
	e^{\gamma T} R^{-5/2} \cos{\!\left( \frac{\pi}{2} \frac{ \log(R/R_1) }{ \log(R_2/R_1) } \right)}
	\,,\quad R_1 \ll R \ll R_2
\end{equation}
where $\gamma$ is given by equation \ref{B2.SSD.real.nonhel.hPr.WKB: eq: gamma soln}.

\subsubsection{Estimates of the turning points}
Recall that equation \ref{B2.SSD.real: eq: dMLdt nonhelical highPr} (which we are analyzing) is valid only for $R \ll 1$.
Further, our act of linearizing in $\etascl$ while substituting for the higher derivatives means that we require $R \gg \sqrt{\etascl}$.
\Textcite[appendix C]{CarSchSch23} assume these scales are good estimates for the turning points, i.e.\@
\begin{equation}
	R_1 \approx \sqrt{\etascl}
	\,,\quad\
	R_2 \approx 1
	\label{B2.SSD.real.nonhel.hPr.WKB: eq: turning point estimates}
\end{equation}
Under these assumptions, we have
\begin{equation}
	\Delta
	\approx
	- \frac{\log\etascl}{2} 
	\scalesAs
	\frac{\log\Rm}{2} 
\end{equation}
where we have used the relation between $\etascl$ and $\Rm$, given in appendix \ref{B2.SSD.real: appendix: Rm and St in terms of tau and eta}.
These estimates suggest that the neglected $\bigO(\Delta^{-2})$ terms in equation \ref{B2.SSD.real.nonhel.hPr.WKB: eq: gamma soln} for $\gamma$ become small when $\left( \log\Rm \right)^2 \gg 1$.

\begin{figure}
	\centering
	\includegraphics{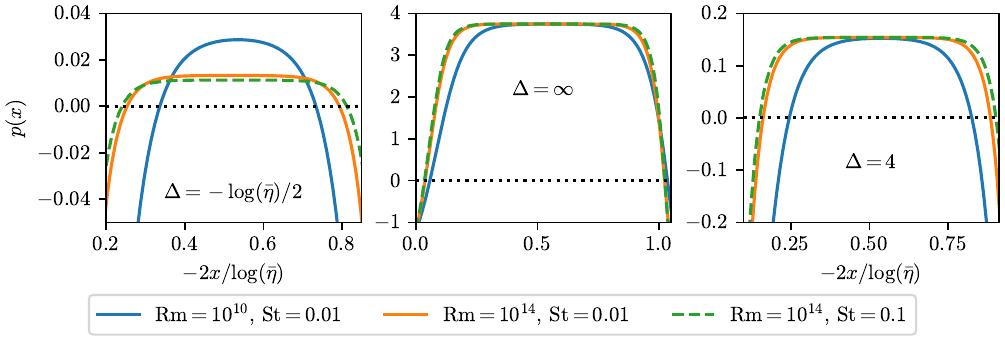}
	\caption{
	The WKB coefficient, $p(x)$, as a function of $x$ for various combinations of $\Rm$ and $\St$.
	Each panel shows a different choice for $\Delta$.
	Recall that the resistive and the integral scales correspond to $-2 x/\log(\etascl) = 0,1$ respectively.
	}
	\label{B2.SSD.real.nonhel.hPr.WKB: fig: p check turning points}
\end{figure}

To convince oneself of these estimates, it is helpful to plot $p(x)$.
However, as noted earlier, the approach we used to derive $p(x)$ above becomes invalid as $x\to\infty$.
To find an expression for $p(x)$ that is valid for arbitrary $x$, we need to go back to the general evolution equation (equation \ref{B2.SSD.real: eq: dMLdt nonhelical with g=0}), and rewrite it in WKB form following the same procedure as we did for its high-$\PrM$ limit:
substitute our chosen form for $E_L(r)$;
assume the solution grows exponentially with time;
use the Landau-Lifshitz approximation to eliminate the third derivative;
and then use equation \ref{B2.SSD.real.nonhel.hPr.WKB: eq: p general} to obtain an expression for $p(x)$ that is valid for all $x$.
To evaluate this expression for given $\Rm$ and $\St$, we use equation \ref{B2.SSD.real.nonhel.hPr.WKB: eq: gamma soln} for the growth rate, and assume the temporal correlation function is exponential.
Since our estimate for the growth rate depends on $\Delta$ (which itself depends on the roots of $p(x)$), the obtained expression for $p(x)$ contains $\Delta$ as a parameter.
Figure \ref{B2.SSD.real.nonhel.hPr.WKB: fig: p check turning points} shows $p(x)$ for different choices of $\Delta$.
We find that regardless of the choice of $\Delta$, the two turning points of $p(x)$ scale like the resistive and the integral scales respectively for high-enough $\Rm$.
This justifies estimating the turning points according to equation \ref{B2.SSD.real.nonhel.hPr.WKB: eq: turning point estimates}.

\subsection{Growth rates for different temporal correlation functions}
\label{section: gamma in terms of St}

Let us now simplify the corrections to the growth rate for the two temporal correlation functions described in table \ref{table: g1 g2 values}.
For exponential temporal correlation, equation \ref{B2.SSD.real: eq: St general} gives us $\tauscl = 2 \St^2 / 3$.
Equation \ref{B2.SSD.real.nonhel.hPr.WKB: eq: gamma soln} for the growth rate (which also assumes a particular form for $E_L(r)$) then becomes
\begin{equation}
	\gamma
	=
	\frac{15}{8} - \frac{375}{32} \, \tauscl
	=
	\gamma_0 \left( 1 - \frac{25 \St^2}{6} \right)
	\approx
	\gamma_0 \left( 1 - 4.2 \St^2 \right)
\end{equation}
On the other hand, for the top hat temporal correlation function, we have $\tauscl = 4 \St^2/ 3$.
The corresponding growth rate can be written as
\begin{equation}
	\gamma
	= 
	\frac{15}{8} - \frac{155}{16} \, \tauscl
	=
	\gamma_0 \left( 1 - \frac{155 \St^2 }{12} \right)
	\approx
	\gamma_0 \left( 1 - 12.9 \St^2 \right)
	\label{B2.SSD.real.nonhel.hPr.comparetcf: eq: gamma top hat}
\end{equation}

\subsection{Comparison with previous work}
In agreement with previous work \citep{Cha97, SchekochihinKulsrud2001, bhat2014fluctuation, CarSchSch23}, we find that the growth rate is reduced by the correlation time of the velocity field being nonzero.\footnote{
In conflict with all these studies, \citet{kleeorin02}, using a renovating flow method, seem to find that the growth rate increases.
This may be related to the issue pointed out in footnote \ref{footnote: Kleeorin mistake}.
}
Further, like \citet{bhat2014fluctuation} and \citet{CarSchSch23}, we find that the spectral slope of the magnetic energy remains unchanged when $\Rm \gg 1$.

%
%
%
%
\begin{figure}
	\centering
	\includegraphics[scale=0.83]{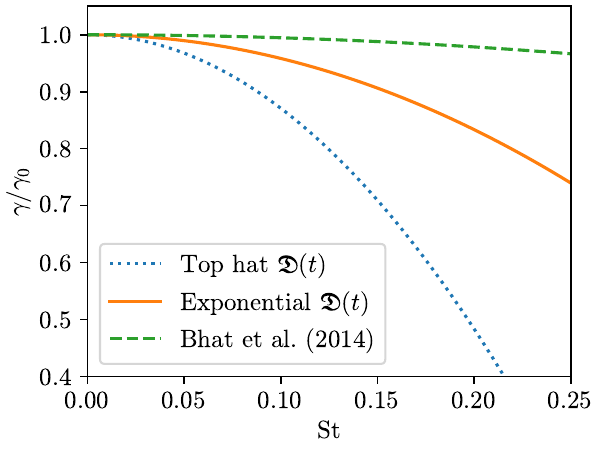}
	\caption{
	Growth rate as a function of $\St$ for two different temporal correlation functions, along with the expression obtained by \textcite[p.~4]{bhat2014fluctuation}.
	}
	\label{B2.SSD.real.nonhel.hPr.comparetcf: fig: gamma vs St}
\end{figure}

Let us now compare the growth rates we found in section \ref{section: gamma in terms of St} with those reported by \textcite{bhat2014fluctuation}.
The velocity field they chose corresponds to the longitudinal correlation function\footnote{
See their eq.~12.
We have accounted for the fact that their definition of the longitudinal correlation function differs from ours by a factor of 2 (see section \ref{B2.SSD.real: compare tau=0 with previous results}).
}
\begin{equation}
	E^\text{BS14}_L(r)
	=
	\frac{A^2 \tau}{6} \left[ 1 + \frac{1}{q^2} \frac{\dh^2}{\dh r^2} \right] j_0(q r)
\end{equation}
where $\tau$ is the renovation time of the velocity field, $q$ is a characteristic wavenumber, $A$ is related to the amplitude of the velocity field, and $j_0$ is the spherical Bessel function of the first kind of order 0 (recall that $j_0(z) = \op{sinc}(z)$).
We identify $q \defn k_f$.
Noting that they define
\begin{equation}
		\eta_t
		\defn
		\frac{ E^\text{BS14}_L(0) }{2}
		=
		\frac{A^2 \tau}{18}
	\,,\quad
		\overline{\tau}^\text{BS14}
		\defn
		\tau \eta_t q^2
		=
		\frac{\tauscl}{2}
		= \frac{2 \St^2}{3}
\end{equation}
Note that for the last equality, we have used $\tauscl = 4 \St^2/ 3$ which corresponds to the temporal correlation function being a top hat.
the growth rate found by them can be written as \citep[p.~4]{bhat2014fluctuation}
\begin{equation}
	\gamma
	=
	\gamma_0 \left( 1 - \frac{15 \St^2}{28} \right)
	\approx
	\gamma_0 \left( 1 - 0.5 \St^2 \right)
\end{equation}
This is shown in figure \ref{B2.SSD.real.nonhel.hPr.comparetcf: fig: gamma vs St}, along with the growth rates we obtained in section \ref{section: gamma in terms of St}.
The suppression of the growth rate is much stronger in our case.

\Textcite[eq.~85]{SchekochihinKulsrud2001} have derived an expression for the growth rates of the single-point moments of the magnetic field at $\PrM \gg 1$ when $\tau_c \ne 0$.
This expression is superficially similar to ours, in that it contains constants parametrizing the temporal correlation properties of the velocity field.
However, they set $\eta = 0$ at the starting of their calculations,\footnote{
See the discussion in their endnote 40.
} while we have taken the limit $\eta\to 0$ only towards the end.
Even when $\tau_c = 0$, this is known to significantly affect the predicted growth rate \parencite[compare eqs.~1.9 and 1.16 of][]{kulsrud92}, and hence we do not expect our $\bigO(\tau_c)$ corrections to match theirs.
Appendix \ref{appendix: relate SchekochihinKulsrud2001} discusses how the quantities we have defined are related to theirs.

\section{Conclusions}
\label{section: conclusions}
By assuming that the velocity field is an incompressible separable Gaussian random field, we have derived the Fourier-space evolution equation for the two-point correlation function of the magnetic field.
Using this equation and further assuming that the velocity field is nonhelical, we have derived the evolution equation for the longitudinal correlation function of the magnetic field ($M_L$) in configuration space, valid for arbitrary $\PrM$ and $\Rm$ (equation \ref{B2.SSD.real: eq: dMLdt nonhelical with g=0}).
Unlike in previous work, setting $\eta = 0$ gives an evolution equation with at most two spatial derivatives of $M_L$.

By choosing an appropriate form for the longitudinal correlation function of the velocity field, we have studied the $\PrM \gg 1$ limit.
In agreement with previous work \citep{Cha97, MasMalBol11, bhat2014fluctuation, CarSchSch23, SchekochihinKulsrud2001}, we have found that the growth rate of the magnetic field decreases when the correlation time is nonzero.
The growth rate is suppressed much more strongly than in the renovating flow model \citep{bhat2014fluctuation, CarSchSch23}.
However, the corrections to the spectral slope of the magnetic field are still negligible when $\Rm \gg 1$.

While our equation \ref{B2.SSD.real: eq: dMLdt nonhelical with g=0} can also be used to study the limit $\PrM \ll 1$
(which has been studied in the white-noise case by, e.g.\@ \citealp{Vin02, ArpHor07, SchSchBov12}; and using a renovating model by \citealp{kleeorin2012growth}),
this seems to be more complicated than the case presented here, and will be described elsewhere.
Further,
it may be interesting to study how the effects of kinetic helicity on the small-scale dynamo \citep{MalBol07, MalBol10} are affected by the correlation time being nonzero.

\software{
	Sympy \citep{sympy2017}.
}

\appendix
\section{Furutsu-Novikov theorem}
\label{appendix: FuruNovi}
Given a functional $R[f]$ of a function $f$, its functional derivative is defined as\footnote{
Some authors \citep[e.g.][]{novikov1965} denote the functional derivative by $\delta R/ (\delta f \d s)$, in order to make its dimensions explicit.
}
\begin{equation}
	\left. \frac{\d^n R[f+\epsilon\chi]}{\d\epsilon^n} \right|_{\epsilon=0} = \int \frac{\delta^n R}{\delta f(s_1)\dots\delta f(s_n)} \, \chi(s_1)\dots\chi(s_n) \,\d s_1\dots\d s_n
	\label{PS: eq: functional derivative definition}
\end{equation}
where $\epsilon$ is a real number, and $\chi(t)$ is an arbitrary test function.
If $f(s)$ is a zero-mean Gaussian random function, $R[f]$ satisfies the Furutsu-Novikov formula \citep{Fur63, novikov1965}:
\begin{equation}
	\meanBr{ f(s) R[f] }
	=
	\int \meanBr{f(s) f(s_1) } \meanBr{\frac{\delta R[f]}{\delta f(s_1)}} \,\d s_1
	\label{eq: FuruNovi}
\end{equation}

Equation \ref{eq: FuruNovi} also holds if $s$ is a collection of spatio-temporal variables and vector indices.

\section{The Fourier-space evolution equation}
\label{appendix: ssdShTerms}

The evolution equation for the two-point single-time correlation of the magnetic field in Fourier space is
\begin{align}
	\begin{split}
		\frac{\dh \meanBr{ \mcB_{ij}^{(\vec{k},t; -\vec{k},t)} } }{\dh t}
		={}&
		- \eta \abs{\vec{k'} }^2 \meanBr{ \mcB_{ij}^{(\vec{k},t; -\vec{k},t)} }
		+ \ssdShTerm{1}_{ij}^{(\vec{k}, t)}
		+ \ssdShTerm{2}_{ij}^{(\vec{k}, t)}
		+ \ssdShTerm{3}_{ij}^{(\vec{k}, t)}
		+ \ssdShTerm{4}_{ij}^{(\vec{k}, t)}
		+ \ssdShTerm{5}_{ij}^{(\vec{k}, t)}
		\\& + \left[ i \exchangeArrow j ;\, \vec{k} \to -\vec{k} \right]
		+ \bigO(\tau_c^2)
	\end{split} \label{B2.SSD.shear: eq: Bij evolution in terms of I* new}
\end{align}
When $\tau_c=0$, only $\ssdShTerm{3}$ and the explicit resistive term are nonzero.
The terms $\ssdShTerm{1}$ and $\ssdShTerm{2}$ come from the second functional derivative of $\mcB$; while $\ssdShTerm{3}$, $\ssdShTerm{4}$, and $\ssdShTerm{5}$ come from Taylor-expanding the $\meanBr{\mcB}$ term that appears during the first application of equation \ref{B2.SSD.shear: eq: recursion relation funDer condensed} to equation \ref{B2.SSD.shear: eq: Bij evolution unsimp condensed}.
Using the convention that $\ssdShTerm{1} = \ssdShTerm{1.1} + \ssdShTerm{1.2} + \dots$ and so on, the terms on the RHS of equation \ref{B2.SSD.shear: eq: Bij evolution in terms of I* new} are\footnote{
To simplify some of these terms, we have assumed $T_{ij}(\vec{k}) = T_{ji}(-\vec{k})$, which would follow from assuming $\tcf(-t) = \tcf(t)$.
As discussed by \textcite{KopIlySir22, KopKisIly22}, the time-asymmetry of the velocity field is closely related to its non-Gaussianity.
Since we have already assumed the velocity field is Gaussian, this additional assumption does not seem very restrictive.
}
\begin{align}
	\begin{split}
		\ssdShTerm{1.1}_{ij}^{(\vec{k},t)}
		={}&
		\tau_c g_1 \meanBr{ \mcB_{iu}^{(\vec{k}, t; -\vec{k}, t)} } \int_{\substack{ \vec{k}^{(1)} }} \iDelLam_{jas}^{(- \vec{k}^{(1)}, -\vec{k} + \vec{k}^{(1)})} \, T_{i_1 a}^{(\vec{k}^{(1)} )}
		\int_{\substack{\vec{p} }} \iDelLam_{smn}^{(\vec{p},-\vec{k} + \vec{k}^{(1)} - \vec{p})} \iDelLam_{n i_1 l}^{(\vec{k}^{(1)},-\vec{k} - \vec{p} )} \iDelLam_{l i_2 u}^{(-\vec{p},-\vec{k} )} \, T_{m i_2}^{(\vec{p} )}
	\end{split}
	\\
	\begin{split}
		\ssdShTerm{1.2}_{ij}^{(\vec{k},t)}
		={}&
		\tau_c g_1 \int_{\substack{ \vec{k}^{(1)}, \vec{q} }} \Dirac^{(\vec{k} - \vec{k}^{(1)} - \vec{q} )} \iDelLam_{jas}^{(- \vec{k}^{(1)}, -\vec{k} + \vec{k}^{(1)} )} \, T_{i_1 a}^{(\vec{k}^{(1)} )} \meanBr{ \mcB_{us}^{(\vec{q}, t; -\vec{q}, t)} }
		\int_{\substack{\vec{p} }} \iDelLam_{imn}^{(\vec{p},\vec{k} - \vec{p})} \iDelLam_{n i_1 l}^{(\vec{k}^{(1)},\vec{k} - \vec{p} - \vec{k}^{(1)})} \iDelLam_{l i_2 u}^{(-\vec{p},\vec{k} - \vec{k}^{(1)})} \, T_{m i_2}^{(\vec{p} )}
	\end{split}
	\\
	\begin{split}
		\ssdShTerm{1.3}_{ij}^{(\vec{k},t)}
		={}&
		\tau_c g_1 \int_{\substack{ \vec{k}^{(1)}, \vec{q}, \vec{p} }} \Dirac^{(\vec{k} - \vec{k}^{(1)} - \vec{p} - \vec{q})} \iDelLam_{jbs}^{(- \vec{k}^{(1)}, -\vec{k} + \vec{k}^{(1)} )} \, T_{i_1 b}^{(\vec{k}^{(1)} )}
		\\&\continuedTerm \times \iDelLam_{sma}^{(-\vec{p},-\vec{k} + \vec{k}^{(1)} + \vec{p})} \iDelLam_{i i_1 l}^{(\vec{k}^{(1)},\vec{k} - \vec{k}^{(1)})} \iDelLam_{l i_2 u}^{(\vec{p},\vec{k} - \vec{k}^{(1)} - \vec{p})} \, T_{i_2 m}^{(\vec{p} )} \meanBr{ \mcB_{ua}^{(\vec{q}, t; -\vec{q}, t)} }
	\end{split}
	\\
	\begin{split}
		\ssdShTerm{1.4}_{ij}^{(\vec{k},t)}
		={}&
		\tau_c g_1 \int_{\substack{\vec{p}, \vec{q} }} \Dirac^{(\vec{k} - \vec{p} - \vec{q})} \iDelLam_{ima}^{(\vec{p},\vec{k} - \vec{p})}  \iDelLam_{l i_2 u}^{(-\vec{p},-\vec{k} + \vec{p})} \, T_{m i_2}^{(\vec{p} )} \meanBr{ \mcB_{au}^{(\vec{q}, t; -\vec{q}, t)} }
		\int_{\substack{ \vec{k}^{(1)} }} \iDelLam_{jbs}^{(- \vec{k}^{(1)}, -\vec{k} + \vec{k}^{(1)} )} \iDelLam_{s i_1 l}^{(\vec{k}^{(1)},-\vec{k} )} \, T_{i_1 b}^{(\vec{k}^{(1)} )}
	\end{split}
	\\
	\begin{split}
		\ssdShTerm{1.5}_{ij}^{(\vec{k},t)}
		={}&
		\tau_c g_1 \int_{\substack{\vec{p}, \vec{q} }} \Dirac^{(\vec{k} - \vec{p} - \vec{q} )} \iDelLam_{i i_2 u}^{(\vec{p},\vec{k} - \vec{p})} \, T_{i_2 m}^{(\vec{p} )} \meanBr{ \mcB_{ua}^{(\vec{q}, t; -\vec{q}, t)} }
		\int_{\substack{ \vec{k}^{(1)} }} \iDelLam_{smn}^{(-\vec{p},-\vec{k} + \vec{k}^{(1)} + \vec{p})} \iDelLam_{n i_1 a}^{(\vec{k}^{(1)},-\vec{k} + \vec{p} )} \iDelLam_{jbs}^{(- \vec{k}^{(1)}, -\vec{k} + \vec{k}^{(1)} )} \, T_{i_1 b}^{(\vec{k}^{(1)} )}
	\end{split}
	\\
	\begin{split}
		\ssdShTerm{1.6}_{ij}^{(\vec{k},t)}
		={}&
		\tau_c g_1 \int_{\substack{ \vec{k}^{(1)}, \vec{p}, \vec{q} }} \Dirac^{(\vec{k} - \vec{p} - \vec{k}^{(1)} - \vec{q} )} \iDelLam_{jbs}^{(- \vec{k}^{(1)}, -\vec{k} + \vec{k}^{(1)} )} \, T_{i_1 b}^{(\vec{k}^{(1)} )}
		\\&\continuedTerm \times \iDelLam_{imn}^{(\vec{p},\vec{k} - \vec{p})} \iDelLam_{n i_1 a}^{(\vec{k}^{(1)},\vec{k} - \vec{p} - \vec{k}^{(1)})} \iDelLam_{s i_2 u}^{(-\vec{p},-\vec{k} + \vec{k}^{(1)} + \vec{p})} \, T_{m i_2}^{(\vec{p} )} \meanBr{ \mcB_{au}^{(\vec{q}, t; -\vec{q}, t)} }
	\end{split}
	\\
	\begin{split}
		\ssdShTerm{1.7}_{ij}^{(\vec{k},t)}
		={}&
		\tau_c g_1 \int_{\substack{ \vec{k}^{(1)}, \vec{q} }} \Dirac^{(\vec{k} - \vec{k}^{(1)} - \vec{q})} \iDelLam_{jbs}^{(- \vec{k}^{(1)}, -\vec{k} + \vec{k}^{(1)} )} \, T_{i_1 b}^{(\vec{k}^{(1)} )} \meanBr{ \mcB_{au}^{(\vec{q}, t; -\vec{q}, t)} }
		\\&\continuedTerm \times \int_{\substack{\vec{p} }} \iDelLam_{sml}^{(\vec{p},-\vec{k} + \vec{k}^{(1)} - \vec{p})} \iDelLam_{i i_1 a}^{(\vec{k}^{(1)},\vec{k} - \vec{k}^{(1)})} \iDelLam_{l i_2 u}^{(-\vec{p}, -\vec{k} + \vec{k}^{(1)} )} \, T_{m i_2}^{(\vec{p} )}
	\end{split}
	\\
	\begin{split}
		\ssdShTerm{1.8}_{ij}^{(\vec{k},t)}
		={}&
		\tau_c g_1 \meanBr{ \mcB_{ua}^{(\vec{k}, t; -\vec{k}, t)} } \int_{\substack{ \vec{k}^{(1)} }} \iDelLam_{jbs}^{(- \vec{k}^{(1)}, -\vec{k} + \vec{k}^{(1)} )} \, T_{i_1 b}^{(\vec{k}^{(1)} )} \int_{\substack{\vec{p} }} \iDelLam_{iml}^{(\vec{p},\vec{k} - \vec{p})} \iDelLam_{s i_1 a}^{(\vec{k}^{(1)},-\vec{k} )} \iDelLam_{l i_2 u}^{(-\vec{p},\vec{k})} \, T_{m i_2}^{(\vec{p} )}
	\end{split}
	\\
	\begin{split}
		\ssdShTerm{2.1}_{ij}^{(\vec{k},t)}
		={}&
		\tau_c g_2 \meanBr{ \mcB_{iu}^{(\vec{k}, t; -\vec{k}, t)} } \int_{\substack{ \vec{k}^{(1)} }} \iDelLam_{jas}^{(- \vec{k}^{(1)}, -\vec{k} + \vec{k}^{(1)} )} \, T_{i_1 a}^{(\vec{k}^{(1)} )}
		\int_{\substack{\vec{p} }} \iDelLam_{smn}^{(\vec{p},-\vec{k} + \vec{k}^{(1)} - \vec{p})} \iDelLam_{n i_2 l}^{(- \vec{p},-\vec{k} + \vec{k}^{(1)} )} \iDelLam_{l i_1 u}^{(\vec{k}^{(1)},-\vec{k} )} \, T_{m i_2}^{(\vec{p} )}
	\end{split}
	\\
	\begin{split}
		\ssdShTerm{2.2}_{ij}^{(\vec{k},t)}
		={}&
		\tau_c g_2 \int_{\substack{ \vec{k}^{(1)}, \vec{q} }} \Dirac^{(\vec{k} - \vec{k}^{(1)} - \vec{q} )} \iDelLam_{jas}^{(- \vec{k}^{(1)}, -\vec{k} + \vec{k}^{(1)} )} \, T_{i_1 a}^{(\vec{k}^{(1)} )} \meanBr{ \mcB_{us}^{(\vec{q}, t; -\vec{q}, t)} }
		\int_{\substack{\vec{p} }} \iDelLam_{imn}^{(\vec{p},\vec{k} - \vec{p})} \iDelLam_{n i_2 l}^{(- \vec{p},\vec{k})} \iDelLam_{l i_1 u}^{(\vec{k}^{(1)},\vec{k} - \vec{k}^{(1)})} \, T_{m i_2}^{(\vec{p} )}
	\end{split}
	\\
	\begin{split}
		\ssdShTerm{2.3}_{ij}^{(\vec{k},t)}
		={}&
		\tau_c g_2 \int_{\substack{ \vec{k}^{(1)}, \vec{p}, \vec{q} }} \Dirac^{(\vec{k} - \vec{p} - \vec{k}^{(1)} - \vec{q} )} \iDelLam_{jbs}^{(- \vec{k}^{(1)}, -\vec{k} + \vec{k}^{(1)} )} \, T_{i_1 b}^{(\vec{k}^{(1)} )} \iDelLam_{sma}^{(-\vec{p},-\vec{k} + \vec{k}^{(1)} + \vec{p})}
		\\&\continuedTerm\times \iDelLam_{i i_2 l}^{(\vec{p},\vec{k} - \vec{p})} \iDelLam_{l i_1 u}^{(\vec{k}^{(1)},\vec{k} - \vec{p} - \vec{k}^{(1)})} \, T_{i_2 m}^{(\vec{p} )} \meanBr{ \mcB_{ua}^{(\vec{q}, t; -\vec{q}, t)} }
	\end{split}
	\\
	\begin{split}
		\ssdShTerm{2.4}_{ij}^{(\vec{k},t)}
		={}&
		\tau_c g_2 \int_{\substack{\vec{p}, \vec{q} }} \Dirac^{(\vec{k} - \vec{p} - \vec{q} )} \iDelLam_{ima}^{(\vec{p},\vec{k} - \vec{p})} \, T_{m i_2}^{(\vec{p} )} \meanBr{ \mcB_{au}^{(\vec{q}, t; -\vec{q}, t)} }
		\int_{\substack{ \vec{k}^{(1)} }} \iDelLam_{s i_2 l}^{(- \vec{p},-\vec{k} + \vec{k}^{(1)} + \vec{p})} \iDelLam_{l i_1 u}^{(\vec{k}^{(1)},-\vec{k} + \vec{p} )} \iDelLam_{jbs}^{(- \vec{k}^{(1)}, -\vec{k} + \vec{k}^{(1)} )} \, T_{i_1 b}^{(\vec{k}^{(1)} )}
	\end{split}
	\\
	\begin{split}
		\ssdShTerm{2.5}_{ij}^{(\vec{k},t)}
		={}&
		\tau_c g_2 \int_{\substack{ \vec{k}^{(1)}, \vec{q} }} \Dirac^{(\vec{k} - \vec{k}^{(1)} - \vec{q} )} \iDelLam_{jbs}^{(- \vec{k}^{(1)}, -\vec{k} + \vec{k}^{(1)} )} \, T_{i_1 b}^{(\vec{k}^{(1)} )} \meanBr{ \mcB_{ua}^{(\vec{q}, t; -\vec{q}, t)} }
		\\&\continuedTerm\times \int_{\substack{\vec{p} }} \iDelLam_{smn}^{(\vec{p},-\vec{k} + \vec{k}^{(1)} - \vec{p})} \iDelLam_{n i_2 a}^{(- \vec{p},-\vec{k} + \vec{k}^{(1)} )} \iDelLam_{i i_1 u}^{(\vec{k}^{(1)},\vec{k} - \vec{k}^{(1)})} \, T_{m i_2}^{(\vec{p} )}
	\end{split}
	\\
	\begin{split}
		\ssdShTerm{2.6}_{ij}^{(\vec{k},t)}
		={}&
		\tau_c g_2 \meanBr{ \mcB_{au}^{(\vec{k}, t; -\vec{k}, t)} } \int_{\substack{ \vec{k}^{(1)} }} \iDelLam_{jbs}^{(- \vec{k}^{(1)}, -\vec{k} + \vec{k}^{(1)} )} \, T_{i_1 b}^{(\vec{k}^{(1)} )} \int_{\substack{\vec{p} }} \iDelLam_{imn}^{(\vec{p},\vec{k} - \vec{p})} \iDelLam_{n i_2 a}^{(- \vec{p},\vec{k})} \iDelLam_{s i_1 u}^{(\vec{k}^{(1)},-\vec{k} )} \, T_{m i_2}^{(\vec{p} )}
	\end{split}
	\\
	\begin{split}
		\ssdShTerm{2.7}_{ij}^{(\vec{k},t)}
		={}&
		\tau_c g_2 \int_{\substack{\vec{p}, \vec{q} }} \Dirac^{(\vec{k} - \vec{p} - \vec{q} )} \iDelLam_{i i_2 a}^{(\vec{p},\vec{k} - \vec{p})} \, T_{i_2 m}^{(\vec{p} )} \meanBr{ \mcB_{au}^{(\vec{q}, t; -\vec{q}, t)} }
		\int_{\substack{ \vec{k}^{(1)} }} \iDelLam_{sml}^{(-\vec{p},-\vec{k} + \vec{k}^{(1)} + \vec{p})} \iDelLam_{l i_1 u}^{(\vec{k}^{(1)},-\vec{k} + \vec{p} )} \iDelLam_{jbs}^{(- \vec{k}^{(1)}, -\vec{k} + \vec{k}^{(1)} )} \, T_{i_1 b}^{(\vec{k}^{(1)} )}
	\end{split}
	\\
	\begin{split}
		\ssdShTerm{2.8}_{ij}^{(\vec{k},t)}
		={}&
		\tau_c g_2 \int_{\substack{ \vec{k}^{(1)}, \vec{p}, \vec{q} }} \Dirac^{(\vec{k} - \vec{p} - \vec{k}^{(1)} - \vec{q} )} \iDelLam_{jbs}^{(- \vec{k}^{(1)}, -\vec{k} + \vec{k}^{(1)} )} \, T_{i_1 b}^{(\vec{k}^{(1)} )}
		\\&\continuedTerm\times \iDelLam_{iml}^{(\vec{p},\vec{k} - \vec{p})} \iDelLam_{s i_2 a}^{(- \vec{p},-\vec{k} + \vec{k}^{(1)} + \vec{p})} \iDelLam_{l i_1 u}^{(\vec{k}^{(1)},\vec{k} - \vec{p} - \vec{k}^{(1)})} \, T_{m i_2}^{(\vec{p} )} \meanBr{ \mcB_{ua}^{(\vec{q}, t; -\vec{q}, t)} }
	\end{split}
	\\
	\begin{split}
		\ssdShTerm{3.1}_{ij}^{(\vec{k},t)}
		={}&
		\meanBr{ \mcB_{in}^{(\vec{k}, t; -\vec{k}, t)} } \int_{\substack{ \vec{k}^{(1)} }} \iDelLam_{jas}^{(- \vec{k}^{(1)}, -\vec{k} + \vec{k}^{(1)} )} \, T_{i_1 a}^{(\vec{k}^{(1)} )} \iDelLam_{s i_1 n}^{(\vec{k}^{(1)},-\vec{k} )}
		\left[ \frac{1}{2} + \frac{\tau_c \eta }{2} \left( - \abs{ \vec{k}^{(1)} - \vec{k} }^2 + \abs{ \vec{k} }^2 \right) \right]
	\end{split}
	\\
	\begin{split}
		\ssdShTerm{3.2}_{ij}^{(\vec{k},t)}
		={}&
		\int_{\substack{ \vec{k}^{(1)}, \vec{q} }} \Dirac^{(\vec{k} - \vec{k}^{(1)} - \vec{q} )} \iDelLam_{jas}^{(- \vec{k}^{(1)}, -\vec{k} + \vec{k}^{(1)} )} \, T_{i_1 a}^{(\vec{k}^{(1)} )} \iDelLam_{i i_1 n}^{(\vec{k}^{(1)},\vec{k} - \vec{k}^{(1)})}
		\left[ \frac{1}{2} + \frac{ \tau_c \eta }{2} \left( - \abs{ \vec{k} }^2 + \abs{ \vec{k} - \vec{k}^{(1)} }^2 \right) \right] \meanBr{ \mcB_{ns}^{(\vec{q}, t; -\vec{q}, t)} }
	\end{split}
	\\
	\begin{split}
		\ssdShTerm{4.1}_{ij}^{(\vec{k},t)}
		={}&
		- \frac{ \tau_c }{4} \meanBr{ \mcB_{ib}^{(\vec{k}, t; -\vec{k}, t)} } \int_{\substack{ \vec{k}^{(1)} }} \iDelLam_{jms}^{(- \vec{k}^{(1)}, -\vec{k} + \vec{k}^{(1)} )} \, T_{i_1 m}^{(\vec{k}^{(1)} )} \iDelLam_{s i_1 n}^{(\vec{k}^{(1)},-\vec{k} )}
		\int_{\substack{ \vec{p} }} \iDelLam_{nla}^{(\vec{p},-\vec{k} - \vec{p})} \, T_{l i_2}^{(\vec{p} )} \iDelLam_{a i_2 b}^{(-\vec{p},-\vec{k} )}
	\end{split}
	\\
	\begin{split}
		\ssdShTerm{4.2}_{ij}^{(\vec{k},t)}
		={}&
		- \frac{ \tau_c }{4} \int_{\substack{ \vec{k}^{(1)}, \vec{q} }} \Dirac^{(\vec{k} - \vec{k}^{(1)} - \vec{q} )} \iDelLam_{jms}^{(- \vec{k}^{(1)}, -\vec{k} + \vec{k}^{(1)} )} \, T_{i_1 m}^{(\vec{k}^{(1)} )} \meanBr{ \mcB_{bs}^{(\vec{q}, t; -\vec{q}, t)} }
		\\&\continuedTerm\times \int_{\substack{ \vec{p} }} \iDelLam_{i i_1 n}^{(\vec{k}^{(1)},\vec{k} - \vec{k}^{(1)})} \iDelLam_{nla}^{(\vec{p},\vec{k} - \vec{k}^{(1)} - \vec{p})} \, T_{l i_2}^{(\vec{p} )} \iDelLam_{a i_2 b}^{(-\vec{p},\vec{k} - \vec{k}^{(1)} )}
	\end{split}
	\\
	\begin{split}
		\ssdShTerm{4.3}_{ij}^{(\vec{k},t)}
		={}&
		- \frac{ \tau_c }{4} \meanBr{ \mcB_{bm}^{(\vec{k}, t; -\vec{k}, t)} } \int_{\substack{ \vec{k}^{(1)} }} \iDelLam_{jns}^{(- \vec{k}^{(1)}, -\vec{k} + \vec{k}^{(1)} )} \, T_{i_1 n}^{(\vec{k}^{(1)} )} \iDelLam_{s i_1 m}^{(\vec{k}^{(1)},-\vec{k} )} \int_{\substack{ \vec{p} }} \iDelLam_{ila}^{(\vec{p},\vec{k} - \vec{p} )} \, T_{l i_2}^{(\vec{p} )} \iDelLam_{a i_2 b}^{(-\vec{p},\vec{k} )}
	\end{split}
	\\
	\begin{split}
		\ssdShTerm{4.4}_{ij}^{(\vec{k},t)}
		={}&
		- \frac{ \tau_c }{4} \int_{\substack{ \vec{k}^{(1)}, \vec{q} }} \Dirac^{(\vec{k} - \vec{k}^{(1)} - \vec{q} )} \iDelLam_{jns}^{(- \vec{k}^{(1)}, -\vec{k} + \vec{k}^{(1)} )} \, T_{i_1 n}^{(\vec{k}^{(1)} )} \iDelLam_{i i_1 m}^{(\vec{k}^{(1)},\vec{k} - \vec{k}^{(1)} )} \meanBr{ \mcB_{mb}^{(\vec{q}, t; -\vec{q}, t)} }
		\\&\continuedTerm\times \int_{\substack{ \vec{p} }} \iDelLam_{sla}^{(\vec{p},-\vec{k} + \vec{k}^{(1)} - \vec{p})} \, T_{l i_2}^{(\vec{p} )} \iDelLam_{a i_2 b}^{(-\vec{p},-\vec{k} + \vec{k}^{(1)} )}
	\end{split}
	\\
	\begin{split}
		\ssdShTerm{5.1}_{ij}^{(\vec{k},t)}
		={}&
		- \frac{ \tau_c }{2} \int_{\substack{ \vec{p}, \vec{q} }} \Dirac^{(\vec{k} - \vec{p} - \vec{q} )} \iDelLam_{nla}^{(-\vec{p},-\vec{k} + \vec{p})} \, T_{i_2 l}^{(\vec{p} )} \iDelLam_{i i_2 b}^{(\vec{p},\vec{k} - \vec{p})} \meanBr{ \mcB_{ba}^{(\vec{q}, t; -\vec{q}, t)} }
		\int_{\substack{ \vec{k}^{(1)} }} \iDelLam_{s i_1 n}^{(\vec{k}^{(1)},-\vec{k} )} \iDelLam_{jms}^{(- \vec{k}^{(1)}, -\vec{k} + \vec{k}^{(1)} )} \, T_{i_1 m}^{(\vec{k}^{(1)} )}
	\end{split}
	\\
	\begin{split}
		\ssdShTerm{5.2}_{ij}^{(\vec{k},t)}
		={}&
		- \frac{\tau_c}{2} \int_{\substack{ \vec{k}^{(1)}, \vec{p}, \vec{q} }} \Dirac^{(\vec{k} - \vec{k}^{(1)} - \vec{p} - \vec{q} )} \iDelLam_{jms}^{(- \vec{k}^{(1)}, -\vec{k} + \vec{k}^{(1)} )} \, T_{i_1 m}^{(\vec{k}^{(1)} )} \iDelLam_{i i_1 n}^{(\vec{k}^{(1)},\vec{k} - \vec{k}^{(1)})}
		\\&\continuedTerm\times \iDelLam_{nla}^{(\vec{p},\vec{k} - \vec{k}^{(1)} - \vec{p})} \, T_{l i_2}^{(\vec{p} )} \iDelLam_{s i_2 b}^{(-\vec{p},-\vec{k} + \vec{k}^{(1)} + \vec{p})} \meanBr{ \mcB_{ab}^{(\vec{q}, t; -\vec{q}, t)} }
	\end{split}
\end{align}

\section{A relation between the coefficients parametrizing the temporal correlation function}
\label{appendix: g_1 + g_2}
Recalling the definitions of $g_1$ and $g_2$ (equations \ref{B2.SSD.shear.tau^1: eq: g1 g2 defn}), we note that
\begin{align}
	\begin{split}
		g_1 + g_2
		={}&
		\frac{1}{\tau_c} \int_{-\infty}^t \d t' \int_{-\infty}^{ t' } \d t_1 \int_{-\infty}^{ t_1 } \d t_2 \, \tcf^{(t-t_1)} \tcf^{(t'-t_2)}
		+ \frac{1}{\tau_c} \int_{-\infty}^t \d t' \int_{-\infty}^{ t' } \d t_2 \int_{-\infty}^{ t_2 } \d t_1 \, \tcf^{(t-t_1)} \tcf^{(t'-t_2)}
	\end{split}
	\\
	\begin{split}
		={}&
		\frac{1}{\tau_c} \int_{-\infty}^t \d t' \int_{-\infty}^{ t' } \d t_1 \int_{-\infty}^{ t' } \d t_2 \, \tcf^{(t-t_1)} \tcf^{(t'-t_2)}
	\end{split}
	\\
	\begin{split}
		={}&
		\frac{1}{2\tau_c} \int_{-\infty}^t \d t' \int_{-\infty}^{ t' } \d t_1 \, \tcf^{(t-t_1)}
	\end{split}
	\\
	\begin{split}
		={}&
		\frac{1}{2\tau_c} \int_{-\infty}^t \d t_1 \int_{t_1}^{ t } \d t' \, \tcf^{(t-t_1)}
	\end{split}
	\\
	\begin{split}
		={}&
		\frac{1}{2\tau_c} \int_{-\infty}^t \d t_1 \left( t - t_1 \right) \tcf^{(t-t_1)}
	\end{split}
	\\
	\begin{split}
		={}&
		\frac{1}{4}
	\end{split}
	\label{B2.SSD.real: eq: g1 + g2}
\end{align}
where we have used the properties of $\tcf(\tau)$, given in equation \ref{B2.SSD.shear.tau^1: eq: wiwj homo separable}.

\section{The relation between \texorpdfstring{$M_L(r)$}{ML(r)} and \texorpdfstring{$M(r)$}{M(r)}}
\label{appendix: M_L M relation lower limit}

Let us denote the lower limit of the integral in equation \ref{B2.SSD.real: eq: M_L M relation} as $r=a$.
Plugging equation \ref{B2.SSD.real: eq: M M_L relation} into the RHS of equation \ref{B2.SSD.real: eq: M_L M relation} and requiring the resulting equation to hold gives us $\lim_{r\to a} r^3 M_L(r) = 0$.
In general, this is only expected to hold for $a=0$ and $a=\infty$.

Alternatively, taking the $r\to 0$ limit of equation \ref{B2.SSD.real: eq: M_L M relation} gives us $\left[ r^3 M_L(r) \right]_{r\to 0} = \int_a^0 r^2 M_{ii} \, \d r$.
Regardless of the value of $a$, equation \ref{B2.SSD.real: eq: M M_L relation} implies
\begin{equation}
	\meanBr{ h_i(\vec{r},t) \, h_i(\vec{r},t) }
	=
	3 M_L(0)
\end{equation}
as long as $\left[ r \, \dh M_L(r) / \dh r \right]_{r=0} = 0$.
This tells us that if the magnetic energy is finite, $M_L(0)$ is also finite.
This means we need $\int_a^0 r^2 M_{ii} \, \d r = 0$.
Recalling that $\int_0^\infty r^2 M_{ii} \, \d r \propto M_{ii}(k=0)$, we see that this condition is satisfied by $a\to\infty$ only if $M_{ii}(k=0) = 0$.
On the other hand, it is trivially satisfied by $a=0$.

As far as the evolution equation for $M_L(r)$ in nonhelical turbulence (equation \ref{B2.SSD.real: eq: dMLdt nonhelical} or \ref{B2.SSD.real: eq: dMLdt nonhelical with g=0}) is concerned, the only effect of choosing $a=\infty$ rather than $a=0$ is a change in the form of the extra terms given in equation \ref{B2.SSD.real: eq: dMLdt extra ML(0) term}.
Elimination of these extra terms is possible in both cases: for $a=\infty$, one requires all the correlation functions to decay faster than any polynomial as $r\to\infty$; for $a=0$, one requires $\left[ \d E_L(r) / \d r \right]_{r=0} = 0$.
Both these requirements seem reasonable.

\section{Fourier identities}
\label{appendix: fourier identities}
We use the following identities (assuming $f(k) \to f(r)$, with the arrow denoting inverse Fourier transformation from $\vec{k}$ to $\vec{r}$; note the second identity holds when $f$ is independent of the direction of $\vec{k}$):
\begin{equation}
		k_i f(k)
		\to
		i \, \frac{\dh f(r)}{\dh r_i}
	\,,\quad
		k^2 f(k)
		\to
		- \frac{1}{r^2} \, \frac{\d}{\d r}{\left( r^2 \, \frac{\d f(r)}{\d r} \right)}
\end{equation}
Further assuming that $f(r=\infty) = 0$, we write\footnote{
It may seem more natural to write
\begin{equation*}
	\frac{ f(k) }{k^2}
	\to
	- \int_0^r \left( \frac{\d r}{r^2}\int_0^r \d r \, r^2 f(r) \right)
\end{equation*}
but this would leave behind extra terms involving $f(0)$ when applied to the Laplacian of $f(r)$.
}
\begin{align}
	\begin{split}
		\frac{ f(k) }{k^2}
		\to{}&
		- \int_r^\infty \left( \frac{\d r}{r^2}\int_r^\infty \d r \, r^2 f(r) \right)
	\end{split} \label{B2.SSD.real.id: eq: inverse lap}
\end{align}

\section{The evolution equation in real space with fourth-order derivatives}
\label{appendix: evol eq g nonzero}

\begin{align}
	\begin{split}
		\frac{\dh M_L}{\dh t}
		={}&
		\frac{1}{r^4} \, \frac{\dh}{\dh r}{\left( \left[ \kappa(r) + \tau_c \, \chi_\tau(r) \right] r^4 \, \frac{\dh M_L}{\dh r} \right)}
		+ \left[ G(r) + \tau_c \, \mathcal{G}_\tau(r) \right] M_L
		+ \tau_c \, A(r) \, \frac{\dh M_L}{\dh r}
		\\& + 2 \tau_c \left( \frac{\eta}{2} \, \frac{\d S_2}{\d r} + \frac{g}{r^4} \, \frac{\d{\left( r^4 S_2^2\right)} }{\d r} \right) \frac{\dh^3 M_L}{\dh r^3}
		+ \tau_c g \, S_2^2(r) \, \frac{\dh^4 M_L}{\dh r^4}
		+ \bigO(\tau_c^2)
	\end{split} \label{B2.SSD.real: eq: dMLdt nonhelical}
\end{align}
where we have defined $g \defn g_1 + g_2 - 1/4$ (in appendix \ref{appendix: g_1 + g_2}, we show that this quantity is actually zero) and
\begin{align}
	\begin{split}
		\chi_\tau(r)
		\defn{}&
		g \left[
			- \frac{11}{4} \left( \frac{\d S_2}{\d r} \right)^{2}
			+ \frac{9}{4} \, \frac{\d^{2}{\left( S_{2}^{2} \right)} }{\d r^{2}}
			+ \frac{13}{r} \, \frac{\d{\left( S_{2}^{2} \right)} }{\d r}
			+ \frac{8}{r^{2}} \, S_{2}^{2}{\left(r \right)}
			\right]
		+ \kappa_\tau(r)
	\end{split}
	\\
	\begin{split}
		\mathcal{G}_\tau(r)
		\defn{}&
		g \Bigg[ 
			\frac{S_{2}{\left(r \right)} S_{2}''''{\left(r \right)}}{2} 
			- \frac{ S_{2}'{\left(r \right)} S_{2}'''{\left(r \right)}}{2} 
			+ \frac{4 S_{2}{\left(r \right)} S_{2}'''{\left(r \right)}}{r} 
			+ \frac{2 S_{2}'{\left(r \right)} S_{2}''{\left(r \right)}}{r} 
			+ \frac{4 S_{2}{\left(r \right)} S_{2}''{\left(r \right)}}{r^{2}} 
			\\&\continuedTerm + \frac{8 \left( S_{2}'{\left(r \right)}\right)^{2} }{r^{2}} 
			- \frac{4 S_{2}{\left(r \right)} S_{2}'{\left(r \right)}}{r^{3}}
			\Bigg]
		+ G_\tau(r)
	\end{split}
	\\
	\begin{split}
		A(r)
		\defn{}&
		- \frac{4 \eta }{r^{5}} \, \frac{\d}{\d r}{\left( r^{4} \, \frac{\d S_{2} }{\d r} \right)}
		- g \left(
			\frac{\d^{3}{\left( S_{2}^{2} \right)} }{\d r^{3}}
			+ \frac{12}{r} \, \frac{\d^{2}{\left( S_{2}^{2} \right)} }{\d r^{2}} 
			+ \frac{36}{r^{2}} \, \frac{\d{\left( S_{2}^{2} \right)}}{\d r}
			+ \frac{24}{r^{3}} \, S_{2}^{2}{\left(r \right)}
		\right)
	\end{split}
\end{align}
The rest of the quantities appearing above are defined in equations \ref{eq: coefficients in evolution equation g=0}.

\section{Dimensionless numbers for a separable velocity correlation function}
\label{B2.SSD.real: appendix: Rm and St in terms of tau and eta}
Given a separable velocity correlation function (see equation \ref{B2.SSD.shear.tau^1: eq: wiwj homo separable}) with longitudinal correlation function $E_L(r)$ and temporal correlation function $\tcf(t)$, equation \ref{eq: urms EL tcf relation} can be written as
\begin{equation}
	u_\text{rms}^2 = 3 \, E_L(0) \, \tcf(0)
\end{equation}
Let us assume the velocity correlation is characterized by a length scale $1/k_f$.
We then write
\begin{equation}
	\St = \tau_c u_\text{rms} k_f
	= \tau_c k_f \sqrt{ 3 \, E_L(0) \, \tcf(0) }
	\label{B2.SSD.real: eq: St general}
\end{equation}
and
\begin{equation}
	\Rm = \frac{ u_\text{rms} }{ \eta k_f }
	= \frac{ \sqrt{ 3 \, E_L(0) \, \tcf(0) } }{\eta k_f}
\end{equation}

For example, exponential temporal correlation (see table \ref{table: g1 g2 values}) would give us
\begin{equation}
		\tau_c
		=
		\frac{2 \St^2 }{3 k_f^2 \, E_L(0)}
	\,,\quad
		\eta
		=
		\frac{3 \, E_L(0) }{ 2 \Rm \St }
\end{equation}
While the numerical factors above depend on the functional form of $\tcf(t)$, we expect $\tau_c \scalesAs \St^2$ and $\eta \scalesAs 1/(\Rm \St)$ to always be valid.
\Citet[eq.~95]{SchekochihinKulsrud2001} and \citet[p.~3]{bhat2014fluctuation} agree that $\tauscl \propto \St^2$.

\section{Relation with a previous calculation of the non-diffusive growth rate of single-point moments}
\label{appendix: relate SchekochihinKulsrud2001}
For the sake of completeness, we note the correspondences between our work and that of \citet{SchekochihinKulsrud2001}.
Their equation 47 for the two-point correlation of the velocity field is
\begin{align}
	\begin{split}
		\meanBr{u_i^{(\vec{x} + \vec{r}, t + \tau)} \, u_j^{(\vec{x}, t)} }
		={}&
		\kappa_0^{(\tau)} \delta_{ij}
		- \frac{1}{2} \, \kappa_2^{(\tau)} \left[ r^2 \delta_{ij} + 2 a \, r_i r_j\right]
		+ \frac{1}{4} \, \kappa_4^{(\tau)} r^2 \left[ r^2 \delta_{ij} + 2 b \, r_i r_j\right]
		+ \dots
	\end{split}
\end{align}
which is more general than the form we have assumed.
Incompressibility can be imposed by choosing
$a = -1/4$,
$b = -2/5$.
Specializing to three spatial dimensions, we obtain (see equation \ref{B2.SSD.real: eq: M_L M relation})
\begin{equation}
	E_L(r) \, \tcf(\tau) = \kappa_0 - \frac{r^2}{4} \, \kappa_2(\tau) + \frac{11 r^4}{140} \, \kappa_4(\tau) + \dots
\end{equation}
On the other hand, we have taken
\begin{equation}
	E_L(r) = E_0 \, e^{-k_f^2 r^2/2}
	= E_0 \left( 1 - \frac{k_f^2 r^2}{2} + \frac{k_f^4 r^4}{8} + \dots\right)
\end{equation}
Comparing both these expressions, we write
\begin{equation}
		\kappa_0(\tau)
		=
		E_0 \, \tcf(\tau)
	\,,\quad
		\kappa_2(\tau)
		=
		2 E_0 k_f^2 \, \tcf(\tau)
	\,,\quad
		\kappa_4(\tau)
		=
		\frac{35}{22} \, E_0 k_f^4 \, \tcf(\tau)
\end{equation}
which gives us $\bar{\kappa}_2 = 2 E_0 k_f^2 $.
Using these, one may calculate the constants $K_1$, $K_2$, and $\tilde{K}_2$ appearing in their equation for the growth rate.
Choosing units where $k_f = E_0 = 1$, their expression for the growth rate of the second moment in incompressible turbulence in three spatial dimensions gives
\begin{equation}
	\gamma_\text{SK01} = 5 + \bigO(\tauscl)
\end{equation}
which is related to our result (equation \ref{B2.SSD.real.nonhel.hPr.WKB: eq: gamma soln})  when $\tauscl = 0$ by
\begin{equation}
	\gamma_0 = \frac{3}{8} \, \gamma_{0, SK01}
\end{equation}
This is exactly the expected relation between the resistive and non-resistive growth rates \parencite[eqs.~1.9, 1.16]{kulsrud92}.

\section{On the validity of the WKB approximation}
\label{section: WKB validity}

\begin{figure}
	\centering
	\includegraphics[scale=0.83]{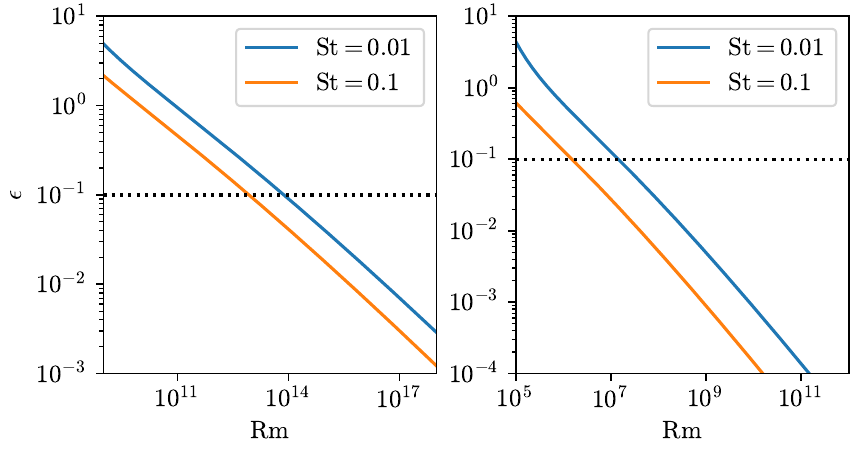}
	\caption{
	Plot of $\epsilon$ (the LHS of equation \ref{B2.SSD.real.nonhel.hPr.WKB.validity: eq: discriminant}) as a function of $\Rm$ for different $\St$.
	The panel on the left shows $\epsilon$ evaluated at $x = -\log(\etascl)/4$ (the geometric mean of the integral and resistive scales);
	the panel on the right shows it at $x = \log\!\left[ \left(1 + 1/\sqrt{\etascl}\right)/2\right]$ (the choice made by \textcite[appendix C.3]{CarSchSch23}).
	}
	\label{B2.SSD.real.nonhel.hPr.WKB: fig: WKB validity}
\end{figure}

As pointed out by \textcite[eq.~C16]{CarSchSch23}, substituting equation \ref{B2.SSD.real.nonhel.hPr.WKB: eq: Theta general} into equation \ref{B2.SSD.real.nonhel.hPr.WKB: eq: WKB form} gives us
\begin{equation}
	\frac{\d^2 \Theta}{\d x^2}
	=
	- p(x) \Theta(x) \left( 
	1
	+ \frac{p''(x) }{4 \left[ p(x) \right]^2}
	- \frac{ 5 \left[ p'(x) \right]^2 }{16 \left[ p(x) \right]^3}
	\right)
\end{equation}
Comparing this with equation \ref{B2.SSD.real.nonhel.hPr.WKB: eq: WKB form}, we find the following consistency condition for the validity of the WKB approximation:
\begin{equation}
	\epsilon \defn
	\abs{
		\frac{p''(x) }{4 \left[ p(x) \right]^2}
		- \frac{ 5 \left[ p'(x) \right]^2 }{16 \left[ p(x) \right]^3}
	} \ll 1
	\label{B2.SSD.real.nonhel.hPr.WKB.validity: eq: discriminant}
\end{equation}
Note that a similar condition has been given by \textcite[eq.~31]{Nor69}.

The LHS of this condition ($\epsilon$) is dependent on $x$.
\Textcite[appendix C.3]{CarSchSch23} evaluate it at the arithmetic mean of the resistive and the integral scales.
However, judging from the form of $p(x)$ (figure \ref{B2.SSD.real.nonhel.hPr.WKB: fig: p check turning points}), we believe it is more appropriate to choose the geometric mean of the resistive and the integral scales (i.e.\@ the arithmetic mean of the logarithms of these scales; recall that $x$ is related to the logarithm of $R$ through equation \ref{B2.SSD.real.nonhel.hPr.WKB: eq: x defn}).
In figure \ref{B2.SSD.real.nonhel.hPr.WKB: fig: WKB validity}, we compare both these choices by using equations \ref{B2.SSD.real.nonhel.hPr.WKB: eq: A0,1,2} and \ref{B2.SSD.real.nonhel.hPr.WKB: eq: p general} to evaluate $p(x)$.
Our choice seems much more conservative.
Further, regardless of the choice made, increasing $\St$ decreases the value of $\Rm$ above which the WKB approximation becomes valid.

\bibliography{refs.bib}
\bibliographystyle{aasjournal}

\end{document}